\documentclass{aastex61}
\usepackage{paralist}
\usepackage{epstopdf}

 \pdfoutput=1 

\received{2017 June 20}
\revised{2017 September 4}
\accepted{2017 September 24}
\submitjournal{ApJS}

\shorttitle{44 GHz methanol masers}
\shortauthors{Rodr\'iguez-Garza et al.}

\begin{document}

\title{A CATALOG OF 44 GHz METHANOL MASERS IN MASSIVE STAR-FORMING REGIONS. \\ IV. The high-mass protostellar object sample}

\correspondingauthor{C. B. Rodr\'iguez-Garza}
\email{ca.rodriguez@crya.unam.mx}

\author{C. B. Rodr\'iguez-Garza}
\affil{Instituto de Radioastronom\'ia y Astrof\'isica, Universidad Nacional Aut\'onoma de M\'exico, Apartado Postal 3-72, Morelia 58089, M\'exico}

\author{S. E. Kurtz}
\affiliation{Instituto de Radioastronom\'ia y Astrof\'isica, Universidad Nacional Aut\'onoma de M\'exico, Apartado Postal 3-72, Morelia 58089, M\'exico}

\author{A. I. G\'omez-Ruiz}
\affiliation{Instituto Nacional de Astrof\'isica, \'Optica y Electr\'onica, Luis E. Erro 1, Tonantzintla, Puebla, C.P. 72840, M\'exico}

\author{P. Hofner}
\altaffiliation{Adjunct Astronomer at the National Radio Astronomy Observatory, 1003 Lopezville Road, Socorro, NM 87801, USA}
\affiliation{Physics Department, New Mexico Tech, 801 Leroy Pl., Socorro, NM 87801, USA}

\author{E. D. Araya}
\affiliation{Physics Department, Western Illinois University, 1 University Circle, Macomb, IL 61455, USA}

\author{S. V. Kalenskii}
\affiliation{Astro Space Center, Lebedev Physical Institute, Profsoyuznaya 84/32, Moscow, 117997, Russia}

\begin{abstract}

We present a survey of 56 massive star-forming regions in the 44 GHz methanol maser transition made with the Karl G. Jansky Very Large Array (VLA);
24 of the 56 fields showed maser emission. The data allow us to demonstrate associations, at arcsecond precision, of the Class I maser emission with outflows, HII regions, and shocks traced by 4.5 micron emission.
We find a total of 83 maser components with linewidths ranging from 0.17 to 3.3 km s$^{-1}$ with a nearly flat distribution and a median value of 1.1 km s$^{-1}$.
The relative velocities of the masers with respect to the systemic velocity of the host clouds range from $-$2.5 to 3.1 km s$^{-1}$ with a distribution peaking near zero.
We also study the correlation between the masers and the so-called extended green objects (EGOs) from the GLIMPSE survey.
Multiple sources in each field are revealed from IR images as well as from centimeter continuum emission from VLA archival data; in the majority of cases the 44 GHz masers are positionally correlated with EGOs which seem to trace the younger sources in the fields. We report a possible instance of a 44 GHz maser associated with a low-mass protostar. If confirmed, this region will be the fifth known star-forming region that hosts Class I masers associated with low-mass protostars. We discuss three plausible cases of maser variability.

\end{abstract}


\section{Introduction \label{sec:intro}}

 Methanol masers are empirically divided into two classes based on their environments and pumping mechanisms (Batrla et al. 1987; Menten et al. 1991). Class I masers are collisionally pumped and sometimes trace dynamically important phenomena such as shocked gas in different astrophysical environments (McEwen et al. 2014; Leurini et al. 2016). Class I masers are often found excited at the interface where molecular outflows or expanding HII regions interact with the quiescent ambient material (Plambeck \& Menten 1990; Araya et al. 2009; Cyganowski et al. 2009; Voronkov et al. 2010), at the interface of supernova remnants with molecular clouds (Pihlstr{\"o}m et al. 2011) and in cloud-cloud collisions (Sobolev et al. 1992). Some of the brightest transitions are at 36 GHz ($4_{-1}\rightarrow 3_0$ E), 44 GHz ($7_0\rightarrow 6_1$ A$^+$), 84 GHz ($5_{-1}\rightarrow 4_0$ E) and 95 GHz ($8_0\rightarrow 7_1$ A$^+$; Cragg et al. 1992). Class II masers are radiatively pumped and occur in close proximity to hot molecular cores, ultracompact HII (UCHII) regions, H$_2$O and OH masers, and embedded IR sources. The most powerful Class II maser transitions are at 6.6 GHz ($5_1 \rightarrow 6_0$ A$^+$) and 12 GHz ($2_0 \rightarrow 3_{-1}$ E; Cragg et al. 1992). Class II methanol masers are exclusively found in high-mass star formation regions (Breen et al. 2013). In contrast, weak Class I methanol masers have been detected toward low-mass protostars (Kalenskii et al. 2010). Although the classification scheme is generally satisfactory, there are exceptions to most of the criteria.


A number of surveys have been made to search for Class I methanol masers. The majority of these were made with single-dish telescopes and mainly at the 44 and 95 GHz transitions (northern hemisphere: Morimoto et al. 1985; Bachiller et al. 1990; Haschick et al. 1990; Kalenskii et al. 1992; Fontani et al. 2010; Gan et al. 2013; Yang et al. 2017; southern hemisphere: Slysh et al. 1994; Val'tts et al. 2000; Ellingsen et al. 2005; Chen et al. 2011). To establish the environments where these masers are excited and their relation to other star formation tracers on arcsecond scales, higher resolution, interferometric surveys are required, but only a few have been carried out to date (northern hemisphere: Kurtz et al. 2004; Cyganowski et al. 2009; G\'omez-Ruiz et al. 2016; McEwen et al. 2016; sourthern hemisphere: Voronkov et al. 2014; Jordan et al. 2015).

 A series of papers has been devoted to VLA studies of 44 GHz Class I masers in high-mass star-forming regions: Kurtz et al. (2004; Paper I), G\'omez et al. (2010; Paper II) and G\'omez-Ruiz et al. (2016; Paper III). Paper I reports the observations of a heterogenous sample of 44 massive star-forming regions spanning from young protostars to compact and ultracompact HII (UCHII) regions. Three of the sources in Paper I could not be imaged and so were re-observed in Paper II to complete the previous survey and provide accurate positions and line parameters. Paper III studied a sub-sample of 69 YSOs selected from the catalog of Molinari et al. (1996). The Paper III contains candidates of high-mass protostellar objects along with more-evolved sources that have already formed UCHII regions; aproximately half of these sources show outflow activity (Zhang et al 2001; 2005).

The present paper is the fourth in the series (Paper IV) which covers the sample of high-mass protostellar objects (HMPOs) reported by Sridharan et al. (2002; hereafter S02). In this paper, we present the data of a 44 GHz VLA survey of Class I methanol masers toward 56 sources of the S02 sample.

In Section \ref{sec:sample}, we describe the sample selection criteria. In Section \ref{sec:obs}, we describe the observational program and data reduction. In Section \ref{sec:res}, we present the results of the survey in tabular form and provide mid-IR and centimenter continuum images. In Section \ref{sec:discuss}, we give a brief discussion and in Section \ref{sec:summ} we give the conclusions. A statistical summary and analysis of Papers I-IV will be presented in a forthcoming paper.

\section{The sample \label{sec:sample}}

The sources observed in the present survey were taken from the sample of 69 HMPOs developed by S02. S02 selected sources north of $-20^\circ$ declination from the IRAS Point Source Catalog (PSC) using four selection criteria: 1) they satisfy the  Wood \& Churchwell (1989) infrared color-color criteria for UCHII regions, 2) they are bright at FIR wavelengths (S$_{60} >$ 90 Jy, S$_{100} >$ 500 Jy), 3) undetected in single-dish radio continuum surveys at 5 GHz to a limit of 25 mJy and 4) detected in the single-dish CS(2$-$1) survey of Bronfman, Nyman, \& May (1996).
The first two criteria favor the selection of massive embedded objects. The third criterion favors young isolated HMPO; i.e., either objects prior to the formation of UCHII regions (which have not yet ionized their surroundings) and/or without more-evolved objects (UCHII regions) in the immediate vicinity. The presence of CS emission indicates that the objects are likely to be embedded within molecular cores.

In a series of papers, Sridharan, Beuther and collaborators (S02, Beuther et al. 2002a, 2002b, 2002c) systematically studied these 69 HMPOs. They characterize the whole sample from mid-IR to centimeter wavelengths; searching for molecular outflows, massive dust cores, ionized gas and maser emission with the aid of single-dish and interferometric observations.

Multiple dust cores with a variety of morphologies were detected in essentially all of the 69 sources through single-dish observations at 1.2 mm (S02, Beuther et al. 2002a). These millimeter peaks are typically offset by several arcseconds from the IRAS position. The brightest core in each field typically has a mass on the order of 10$^2 - 10^3$ M$_\odot$; the H$_2$ column densities are typically a few times 10$^{23}$ cm$^{-2}$, which corresponds to a few hundred magnitudes of visual extinction; average gas densities are around 10$^5$ cm$^{-3}$.

Dense gas in these regions was detected toward 40 of the 69 sources by single-dish and interferometric observations of the NH$_3$ (1,1) and (2,2) inversion lines (S02; Lu et al. 2014). The average rotation temperature and deconvolved linewidth are 18 K and 1.1 km s$^{-1}$, respectively (Lu et al. 2014). Three morphology types were identified: filamentary, centrally-peaked and diffuse. The ammonia emission coincides with the dust continuum emission peaks at 1.2 mm reported by Beuther et al. (2002a).

Eighty-five percent of the sample shows evidence of massive and energetic bipolar outflows with parsec-scale sizes based on single-dish observations of the CO (2$-$1) line (S02, Beuther et al. 2002b). In most cases, the outflows are centered on the mm peaks, which likely harbor the most massive and youngest proto-stars.

In spite of the non-detection in single-dish centimeter continuum surveys at a level of 25 mJy (Gregory \& Condon 1991; Griffith et al. 1994; Wright et al. 1994), more sensitive interferometric studies revealed such emission in 40 of the 69 sources at a detection limit of 1 mJy (S02). No free-free emission above 1 mJy was found in the remaining 29 sources (S02). However, deeper VLA observations toward 20 of the 29 non-detections found thermal free-free emission at the $\mu$Jy level ($\sim 3-50$ $\mu$Jy) in all sources (Zapata et al. 2006; Rosero et al. 2016). Thus, 60 of the 69 sources have centimeter continuum emission.

Maser emission is also associated with these objects. The single-dish studies of S02 revealed 26 fields with 6.6 GHz Class II CH$_3$OH masers and 29 fields with H$_2$O masers; Beuther et al. (2002c) report interferometric positions for some of these. OH masers are also associated with only three sources of the sample (Forster et al. 1999; Argon et al. 2000). 

A summary of these various star-formation indicators is given in Table \ref{table:refs} for those sources with 44 GHz maser detections.

\startlongtable
\begin{deluxetable}{lccccclc}
\tabletypesize{\footnotesize}
\tablecolumns{8} 
\tablewidth{0pt}
\tablecaption{Characteristic parameters of the 24 HMPOs associated with 44 GHz methanol masers.\label{table:refs}}
\tablehead{
\colhead{IRAS} & \colhead{CH$_3$OH} &  \colhead{H$_2$O} &  \colhead{OH} & \colhead{CO} &  \multicolumn{2}{c}{Radio (3.6 cm; Figure \ref{irac})\tablenotemark{*}}  &  \colhead{T$_{\rm{rot}}$(NH$_3$)} \\
\cline{6-7} 
\colhead{Name}           & 
\colhead{6.6 GHz}          & 
\colhead{22 GHz}         &
\colhead{1665 MHz}               &
\colhead{Outflow/Wings}  &
\colhead{$\sigma$($\mu$Jy beam$^{-1}$)}       &
\colhead{Contours}       &
\colhead{(K)}
 }
\colnumbers
\startdata
05358$+$3543 & 162                & 45      & 1.72\tablenotemark{b}       & +        &  \nodata               & \nodata                            & 18                   \\
18089$-$1732 & 54                 & 75      & 4.16\tablenotemark{b}       & +        &  40\tablenotemark{d}   & $-$4, 4, 6, 8, 10                  & 38                   \\
18102$-$1800 & 8.8                & \nodata & \nodata                     & +        &  77\tablenotemark{h}   & $-3$, 3, 6, 9                      & 15                    \\
18151$-$1208 & 50                 & 0.8     & \nodata                     & +        &  6\tablenotemark{e}    & $-$2, 3, 9, 15, 25                 & 17                    \\
18182$-$1433 & 24                 & 18      & 2.2\tablenotemark{c}        & +        &  40\tablenotemark{d}   & $-$4, 4, 6, 8, 10                  & 20                    \\
18247$-$1147 & 1.6                & \nodata & \nodata                     & +        &  500\tablenotemark{i}  & $-$3, 3, 15, 30, 60                & \nodata                \\
18264$-$1152 & 3.8                & 50      & \nodata                     & +        &  40\tablenotemark{d}   & $-$4, 4, 6, 8                      & 18                     \\
18290$-$0924 & 15.1               & 4       & \nodata                     & +        &  70\tablenotemark{i}   & $-$3, 3, 4, 5, 6                   & 20                    \\
18306$-$0835 & \nodata            & 0.7     & \nodata                     & +        &  520\tablenotemark{i}  & $-$3, 3, 9, 15, 30, 45             & \nodata                  \\
18308$-$0841 & \nodata            & 1.5     & \nodata                     & +        &  45\tablenotemark{d}   & $-$5, 5, 10, 20, 30, 60, 90        & 18                      \\
18310$-$0825 & 12\tablenotemark{a} & \nodata & \nodata                    & +        &  70                    & $-$4, 4, 8, 12, 16                 & 18                      \\
18337$-$0743 & \nodata            & \nodata & \nodata                     & $-$      &  7\tablenotemark{e}    & $-$2, 3, 5, 9, 12                  & 17                      \\
18345$-$0641 & 5.4                & 3       & \nodata                     & +        &  7\tablenotemark{e}    & $-$1.2, 3, 5, 6.5                  & 16                       \\
18488$+$0000 & 16.9               & 1       & \nodata                     & +        &  170\tablenotemark{h}  & $-$4, 4, 8, 12                     & 20                        \\
18517$+$0437 & 279                & 45.3    & \nodata                     & +        &  4.5\tablenotemark{e}  & $-2$, 3, 5, 7, 11                     & 8$-$31\tablenotemark{g}   \\
18521$+$0134 & 1.3                & \nodata & \nodata                     & $-$      &  5\tablenotemark{e}    & $-$2, 3, 7, 9, 12                  & 8$-$37\tablenotemark{g}       \\
18530$+$0215 & \nodata            & \nodata & \nodata                     & +        &  330                   & $-$4, 4, 12, 20, 30, 40, 50        & 16                         \\
18553$+$0414 & \nodata            & 50      & \nodata                     & +        &  4.5\tablenotemark{e}  & $-$1.5, 3, 9, 21, 31               & 8$-$27\tablenotemark{g}     \\
18566$+$0408 & 7.2                & 3       & \nodata                     & +        &  4\tablenotemark{e}    & $-$2, 3, 5, 7, 9, 13, 20, 29       & 15                             \\
20126$+$4104 & 36                 & 15      & \nodata                     & +        &  16\tablenotemark{f}   & $-$3, 3, 5, 6, 7                   & 23                            \\
20293$+$3952 & \nodata            & 100     & \nodata                     & +        &  70                    & $-$3, 3, 4, 5, 6                   & 15                          \\
23033$+$5951 & \nodata            & 4       & \nodata                     & +        &  60                    & $-$4, 5, 6, 7, 8, 9                & 20                          \\
23139$+$5939 & 2.6                & 400     & \nodata                     & +        &  60                    & $-$4, 5, 6, 7, 8, 9, 10            & 8$-$31\tablenotemark{g}     \\
23151$+$5912 & \nodata            & 60      & \nodata                     & +        &  \nodata               & \nodata                            & \nodata                      \\         
\enddata
\tablecomments{The numbers in columns 2, 3, and 4 are the maser peak flux densities in units of Jansky. Except when indicated, the maser fluxes, outflow detections, 3.6 cm emission and rotational temperatures are taken from S02. The plus and minus signs are detections and nondetections, respectively.}
\tablenotetext{a}{Szymczak et al. (2000).}
\tablenotetext{b}{Argon et al. (2000).}
\tablenotetext{c}{Forster et al. (1999).}
\tablenotetext{d}{C-configuration VLA archival observations from Zapata et al. (2006).}
\tablenotetext{e}{A-configuration VLA archival observations made at 6 cm from Rosero et al. (2016).}
\tablenotetext{f}{B-configuration VLA archival data from Hofner et al. (1999).}
\tablenotetext{g}{Lu et al. (2014).}
\tablenotetext{h}{We applied the UVTAPER parameter with a value of 100 k$\lambda$.}
\tablenotetext{i}{We applied the UVTAPER parameter with a value of 80 k$\lambda$.}
\tablenotetext{*}{The parameters reported here correspond to the rms and contour levels (in multiples of the rms) from Figure \ref{irac}.}
\end{deluxetable}

\section{Observations and data reduction \label{sec:obs}}

The observations were made with the NRAO\footnote{The National Radio Astronomy Observatory (NRAO) is operated by Associated Universities, Inc., under a cooperative agreement with National Science Foundation.} VLA in D configuration (project code AK0699) during seven observing runs on 2008 August 26 and 31, and 2008 September 05, 07, 08, 12 and 14. A summary of the flux and phase calibrators used in each observing period is given in Table \ref{table:calibrator}. The observations on September 12 were divided in time ranges separated by about an hour. The first time range suffered from poor weather conditions; this is reflected in the large channel map rms reported in Table \ref{table:sources}. Ten sources were observed in this group and no masers were detected in any of them.

The data were observed in Q band (0.7 cm) using the dual IF mode and the fast-switching method. One IF was centered on the methanol maser transition (7$_0 \rightarrow 6_1$ A$^+$) with a rest frequency of 44069.43 MHz. A 3.125 MHz bandwidth was used, divided in 127 channels of 24.4 kHz each, providing a spectral resolution of 0.16 km s$^{-1}$ and a velocity coverage of 21 km s$^{-1}$.

\begin{deluxetable}{lllcl}
\tablecolumns{5} 
\tablewidth{0pt}
\tablecaption{Observational program summary \label{table:calibrator}}
\tablehead{
\colhead{Observation}                   & 
\colhead{Phase}                         & 
\colhead{Flux density}                  &
\colhead{Calibrator\tablenotemark{$\dagger$}} &
\colhead{Sources calibrated}            \\
\colhead{date}                          & 
\colhead{calibrator}                    & 
\colhead{(Jy)}                          &
\colhead{code}                          &
\colhead{}                              
}  
\colnumbers
\startdata
2008 Aug 26  & J0555$+$398   & 2.06 $\pm$ 0.06 & A & 05358+3543, 05490+2658                                        \\
             & J0539$+$145   & 0.34 $\pm$ 0.01 & A & 05553+1631                                                    \\
2008 Aug 31  & J1733$-$130   & 4.42 $\pm$ 0.15 & B & 18089$-$1732, 18090$-$1832, 18102$-$1800, 18151$-$1208, 18159$-$1550,  \\
             &               &                 &   & 18182$-$1433, 18223$-$1243                           \\                           
2008 Sep 05  & J2015$+$371   & 3.10 $\pm$ 0.08 & C & 20051+3435, 20126+4104, 20205+3948                              \\
             & J2023$+$318   & 1.04 $\pm$ 0.02 & A & 20081+2720                                                       \\
             & J2007$+$404   & 1.74 $\pm$ 0.04 & B & 20216+4107, 20293+3952, 20319+3958, 20332+4124, 20343+4129              \\
2008 Sep 07  & J1832$-$105   & 0.57 $\pm$ 0.01 & C & 18247$-$1147, 18264$-$1152, 18272$-$1217, 18290$-$0924, 18306$-$0835,          \\
             &               &                 &   & 18308$-$0841                                              \\
2008 Sep 08  & J1832$-$105   & 0.59 $\pm$ 0.01 & C & 18310$-$0825, 18337$-$0743, 18345$-$0641, 18348$-$0616, 18372$-$0541          \\
2008 Sep 12  & J2022$+$616   & 1.02 $\pm$ 0.04 & B & 22134+5834                                                       \\
             & J0014$+$612   & 0.52 $\pm$ 0.02 & C & 22551+6221, 22570+5912, 23033+5951, 23139+5939, 23151+5912     \\
             &               &                 &   & 23545+6508                                             \\
2008 Sep 12\tablenotemark{*} & J1851$+$005     & 0.90 $\pm$ 0.05 & C & 18385$-$0512, 18426$-$0204, 18431$-$0312, 18437$-$0216, 18440$-$0148,   \\
             &               &                 & &  18445$-$0222, 18447$-$0229, 18454$-$0158, 18454$-$0136, 18460$-$0307                \\
2008 Sep 14  & J1851$+$005   & 0.81 $\pm$ 0.01 & C & 18470$-$0044, 18472$-$0022, 18488+0000, 18517+0437, 18521+0134,           \\
             &               &                 &   & 18530+0215,  18540+0220, 18553+0414, 18566+0408                 \\
\enddata
\tablenotetext{\dagger}{The codes A, B, and C correspond to positional accuracy of $<$ 0\rlap.{$''$}002, 0\rlap.{$''$}002$-$0\rlap.{$''$}01, and 0\rlap.{$''$}01$-$0\rlap.{$''$}15, respectively.}
\tablenotetext{*}{Observing run with large channel map rms and no maser detection (see Table \ref{table:masers}).}
\end{deluxetable}

We performed a standard calibration with the AIPS software package. We used channel-0 data (comprised of the central 75\% of the bandpass) to calibrate the complex instrumental gain of each antenna and the results were copied to the line data set. The flag and calibration tables were applied to the line data and then individual sources were imaged. No bandpass calibration was performed. At the time of the observations, Doppler tracking was not supported owing to the VLA to EVLA conversion; Doppler corrections were made with the CVEL task.

For each field we made a dirty data cube to identify the channel of the brightest maser component. We made a clean image of this peak channel, setting a clean box around the maser emission. If the signal-to-noise ratio was high enough, the cleaned channel was used as a model to self-calibrate the data. A first iteration was performed in phase with short solution intervals. When appropriate, we did a second iteration in amplitude and phase with longer solution intervals. Finally, we made a clean data cube setting boxes around all the maser components identified. In this step we fixed the parameter FLUX to clean to a 5$\sigma$ level, where $\sigma$ is the theoretical noise for each field.

Table \ref{table:sources} shows the general information of each field. We list the associated IRAS source name, the pointing center in J2000 coordinates, the central velocity of the bandpass (set to the systemic velocity reported by S02), and the synthesized beam size of the image. The 1$\sigma$ rms noise is given in column 7; it was obtained from the central portion of the final data cube, but excluding channels with maser emission. In the last column, we report the number of maser components detected.

\startlongtable
\begin{deluxetable}{cccccccc}
\tabletypesize{\footnotesize}
\tablecolumns{8} 
\tablewidth{0pt}
\tablecaption{Observed source list}
\tablehead{
\colhead{Source} & \multicolumn{2}{c}{J2000 Pointing Center} & \colhead{Central Velocity} & \multicolumn{2}{c}{Synthesized Beam}
    & \colhead{Channel Map rms} & \colhead{Maser}\\
\cline{2-3} \cline{5-6}
\colhead{IRAS name}            & 
\colhead{$\alpha$($^{\rm h~ m~ s}$)}   & 
\colhead{$\delta$($^\circ$  $'~ ''$)}   & 
\colhead{(km s$^{-1})$}      &
\colhead{(arcsec)}  & 
\colhead{(deg)}   & 
\colhead{(mJy beam$^{-1})$}   & 
\colhead{Detection}  
 } 
\startdata
05358$+$3543 & 05 39 10.4 & $+$35 45 19 & $-$17.6 & 1.69$\times$1.48  & $-$57.0  & 39 & 3\\
05490$+$2658 & 05 52 12.9 & $+$26 59 33 & $+$0.8  & 1.57$\times$1.43  & $-$83.5  & 51 & 0\\
05553$+$1631 & 05 58 13.9 & $+$16 32 00 & $+$5.7  & 1.67$\times$1.48  & $-$49.9  & 51 & 0\\  
18089$-$1732 & 18 11 51.3  & $-$17 31 29  & $+$33.8 & 2.16$\times$1.35  & $-$4.1  & 76 & 7 \\
18090$-$1832 & 18 12 01.9  & $-$18 31 56  & $+$109.8 & 2.21$\times$1.33 & $-$0.3  & 79 & 0 \\
18102$-$1800 & 18 13 12.2   & $-$17 59 35 & $+$21.1  & 2.27$\times$1.35 & $+$2.1  & 85 & 6 \\
18151$-$1208 & 18 17 57.1   & $-$12 07 22 & $+$32.8 & 2.05$\times$1.42  & $+$1.9  & 66 & 4\tablenotemark{a} \\
18159$-$1550 & 18 18 47.3   & $-$15 48 58 & $+$59.9 & 2.25$\times$1.43  & $+$3.9  & 66 & 0 \\
18182$-$1433 &  18 21 07.9  & $-$14 31 53 &  $+$59.1 & 2.18$\times$1.47 & $+$2.2  & 69 & 4 \\
18223$-$1243 & 18 25 10.9  &  $-$12 42 17 & $+$45.5 & 2.12$\times$1.45  & $+$7.9  & 78 & 0 \\
18247$-$1147 & 18 27 31.1  & $-$11 45 56  & $+$121.7& 2.14$\times$1.47 & $-$4.9 & 43 & 1 \\
18264$-$1152 & 18 29 14.3  & $-$11 50 26  & $+$43.6 & 2.13$\times$1.47 & $-$0.1 & 41 & 10 \\  
18272$-$1217 & 18 30 02.7  & $-$12 15 27  & $+$34.0 & 2.13$\times$1.47 & $+$3.9 & 44 & 0 \\
18290$-$0924 & 18 31 44.8  & $-$09 22 09  & $+$84.3 & 2.08$\times$1.56 & $-$6.5 & 72 & 3  \\
18306$-$0835 & 18 33 21.8  & $-$08 33 38  & $+$76.8 & 2.13$\times$1.48 & $+$14.2& 36 & 3  \\  
18308$-$0841 & 18 33 31.9  & $-$08 39 17  & $+$77.1 & 2.05$\times$1.51 & $+$12.8& 44 & 5 \\
18310$-$0825 & 18 33 47.2 & $-$08 23 35 & $+$84.4 & 1.99$\times$1.52 & $-$10.8 & 45 & 1 \\
18337$-$0743 & 18 36 29.0 & $-$07 40 33 & $+$57.9 & 1.95$\times$1.53 & $-$12.4 & 44 & 2 \\  
18345$-$0641 & 18 37 16.8 & $-$06 38 32 & $+$95.9 & 1.89$\times$1.54 & $-$10.4 & 41 & 8 \\
18348$-$0616 & 18 37 29.0 & $-$06 14 15 & $+$109.5& 1.87$\times$1.52 & $-$3.5  & 42 & 0 \\
18372$-$0541 & 18 39 56.0 & $-$05 38 49 & $+$23.6 & 1.90$\times$1.53 & $+$1.0  & 50 & 0 \\
18385$-$0512 & 18 41 12.0 & $-$05 09 06 & $+$26.0 & -- & -- & -- & \tablenotemark{bc} \\
18426$-$0204 & 18 45 12.8 & $-$02 01 12 & $+$15.0 & -- & -- & -- & \tablenotemark{bc} \\
18431$-$0312 & 18 45 46.9 & $-$03 09 24 & $+$105.2 & 1.77$\times$1.34 & $+$20.9 & 404 & 0\tablenotemark{c}  \\
18437$-$0216 & 18 46 22.7 & $-$02 13 24 & $+$110.8 & 1.84$\times$1.45 & $+$29.6 & 191 & 0\tablenotemark{c} \\
18440$-$0148 & 18 46 36.3 & $-$01 45 23 & $+$97.6  & 1.87$\times$1.52 & $+$20.9 & 134 & 0\tablenotemark{c} \\
18445$-$0222 & 18 47 10.8 & $-$02 19 06 & $+$86.8  & 1.94$\times$1.51 & $+$29.2 & 113 & 0\tablenotemark{c} \\
18447$-$0229 & 18 47 23.7 & $-$02 25 55 & $+$102.6 & 1.97$\times$1.51 & $+$29.3 & 110 & 0\tablenotemark{c} \\
18454$-$0158 & 18 48 01.3 & $-$01 54 49 & $+$52.6  & 2.03$\times$1.49 & $+$35.4 & 112 & 0\tablenotemark{c}\\
18454$-$0136 & 18 48 03.7 & $-$01 33 23 & $+$38.9  & 1.98$\times$1.52 & $+$36.5 & 110 & 0\tablenotemark{c} \\
18460$-$0307 & 18 48 39.2 & $-$03 03 53 & $+$83.7  & 2.10$\times$1.50 & $+$36.9 & 105 & 0\tablenotemark{c} \\
18470$-$0044 & 18 49 36.7 & $-$00 41 05 & $+$96.5 & 1.75$\times$1.49 & $+$14.1 & 50 & 0 \\
18472$-$0022 & 18 49 50.7 & $-$00 19 09 & $+$49.0 & 1.75$\times$1.50 & $+$19.0 & 48 & 0 \\
18488$+$0000 & 18 51 24.8 & $+$00 04 19 & $+$82.7 & 1.76$\times$1.50 & $+$18.3 & 46 & 1 \\
18517$+$0437 & 18 54 13.8 & $+$04 41 32 & $+$43.9 & 1.71$\times$1.52 & $+$17.6 & 48 & 2\tablenotemark{d} \\
18521$+$0134 & 18 54 40.8 & $+$01 38 02 & $+$76.0 & 1.76$\times$1.52 & $+$33.0 & 45 & 2 \\
18530$+$0215 & 18 55 34.2 & $+$02 19 08 & $+$77.7 & 1.81$\times$1.55 & $+$21.7 & 46 & 2 \\
18540$+$0220 & 18 56 35.6 & $+$02 24 54 & $+$49.6 & 1.83$\times$1.55 & $+$29.1 & 48 & 0 \\
18553$+$0414 & 18 57 52.9 & $+$04 18 06 & $+$10.0 & 1.77$\times$1.54 & $+$37.5 & 50 & 1 \\
18566$+$0408 & 18 59 09.9 & $+$04 12 14 & $+$85.2 & 1.70$\times$1.53 & $+$45.6 & 46 & 5 \\
20051$+$3435 & 20 07 03.8  &  $+$34 44 35 & $+$11.6 & 1.69$\times$1.47 & $-$72.3 & 43 & 0\\
20081$+$2720 & 20 10 11.5  & $+$27 29 06  & $+$5.7  & 1.83$\times$1.54 & $-$89.1 & 47 & 0 \\
20126$+$4104 & 20 14 26.0  & $+$41 13 32  & $-$3.8  & 1.72$\times$1.44 & $-$67.4 & 77 & 5\tablenotemark{de}\\ 
20205$+$3948 & 20 22 21.9  & $+$39 58 05  & $-$1.7  & 1.73$\times$1.42 & $-$66.5 & 42 & 0 \\   
20216$+$4107 & 20 23 23.8  & $+$41 17 40  & $-$2.0  & 1.79$\times$1.43 & $-$72.3 & 43 & 0 \\
20293$+$3952 & 20 31 10.7  & $+$40 03 10  & $+$6.3  & 1.80$\times$1.43 & $-$78.2 & 42 & 3 \\
20319$+$3958 & 20 33 49.3  & $+$40 08 45  & $+$8.8  & 1.82$\times$1.42 & $-$72.9 & 46 & 0 \\
20332$+$4124 & 20 35 00.5  & $+$41 34 48  & $-$2.0  & 1.86$\times$1.42 & $-$79.2 & 43 & 0 \\
20343$+$4129 & 20 36 07.1  & $+$41 40 01  & $+$11.5 & 1.88$\times$1.43 & $-$79.5 & 46 & 0 \\
22134$+$5834 & 22 15 09.1 & $+$58 49 09 & $-$18.3 & 1.65$\times$1.28 & $-$15.8 & 62 & 0 \\
22551$+$6221 & 22 57 05.2 & $+$62 37 44 & $-$13.4 & 1.80$\times$1.26 & $-$11.6 & 78 & 0 \\
22570$+$5912 & 22 59 06.5 & $+$59 28 28 & $-$46.7 & 1.70$\times$1.29 & $-$15.6 & 76 & 0 \\
23033$+$5951 & 23 05 25.7 & $+$60 08 08 & $-$53.1 & 1.74$\times$1.28 & $-$19.3 & 78 & 2 \\ 
23139$+$5939 & 23 16 09.3 & $+$59 55 23 & $-$44.7 & 1.73$\times$1.30 & $-$17.4 & 85 & 1 \\
23151$+$5912 & 23 17 21.0 & $+$59 28 49 & $-$54.4 & 1.62$\times$1.27 & $-$19.4 & 83 & 2 \\ 
23545$+$6508 & 23 57 05.2 & $+$65 25 11 & $-$18.4 & 1.78$\times$1.34 & $-$10.4 & 85 & 0 \\ 
\enddata
\tablenotetext{a}{The channels in the velocity range of 29.4 to 31.6 km s$^{-1}$ presented an rms noise about six times larger than in the rest of the data cube.
                  Our detection limit in this velocity range is $\sim$ 2 Jy.}
\tablenotetext{b}{This source was not calibrated because of low SNR of the phase calibrator.}
\tablenotetext{c}{Observations made under bad weather conditions on 2008 September 12 in the timerange from 03:15:00 to 05:14:20. The null maser detection is inconclusive.}
\tablenotetext{d}{This source was also observed in Paper II.}
\tablenotetext{e}{This source was also observed in Paper I.}
\label{table:sources}
\end{deluxetable}

For the pointing center of each field we used the position given by S02. We caution that these positions (given in their Table 1 and also our Table \ref{table:sources}) in general are not the same as the IRAS PSC positions. The offsets from the IRAS positions range from 0\rlap.{$''$}10 to $33''$; the biggest separation is for IRAS 18306$-$0835. In some cases, this positional difference may be due to a truncation in the coordinate values (T.K. Sridharan, private communication). In Figure \ref{irac} we show the IRAS error ellipses centered on the PSC position given by the SIMBAD astronomical database\footnote{This research has made use of the SIMBAD database, operated at CDS, Strasbourg, France.}. We also show the primary beam of these observations ($\sim 1'$) centered on the positions reported by S02. Except for 18306$-$0835, the IRAS error ellipses fall almost entirely within the VLA primary beam.

For the source 18151$-$1208, the channels in the velocity range of 29.4 km s$^{-1}$ to 31.6 km s$^{-1}$ presented an rms noise about six times larger than in the rest of the data cube. Our detection limit for masers in this velocity range is about 2 Jy.

\begin{figure*}[h!]
\begin{center}
\begin{tabular}{ll}
\includegraphics[width=71mm]{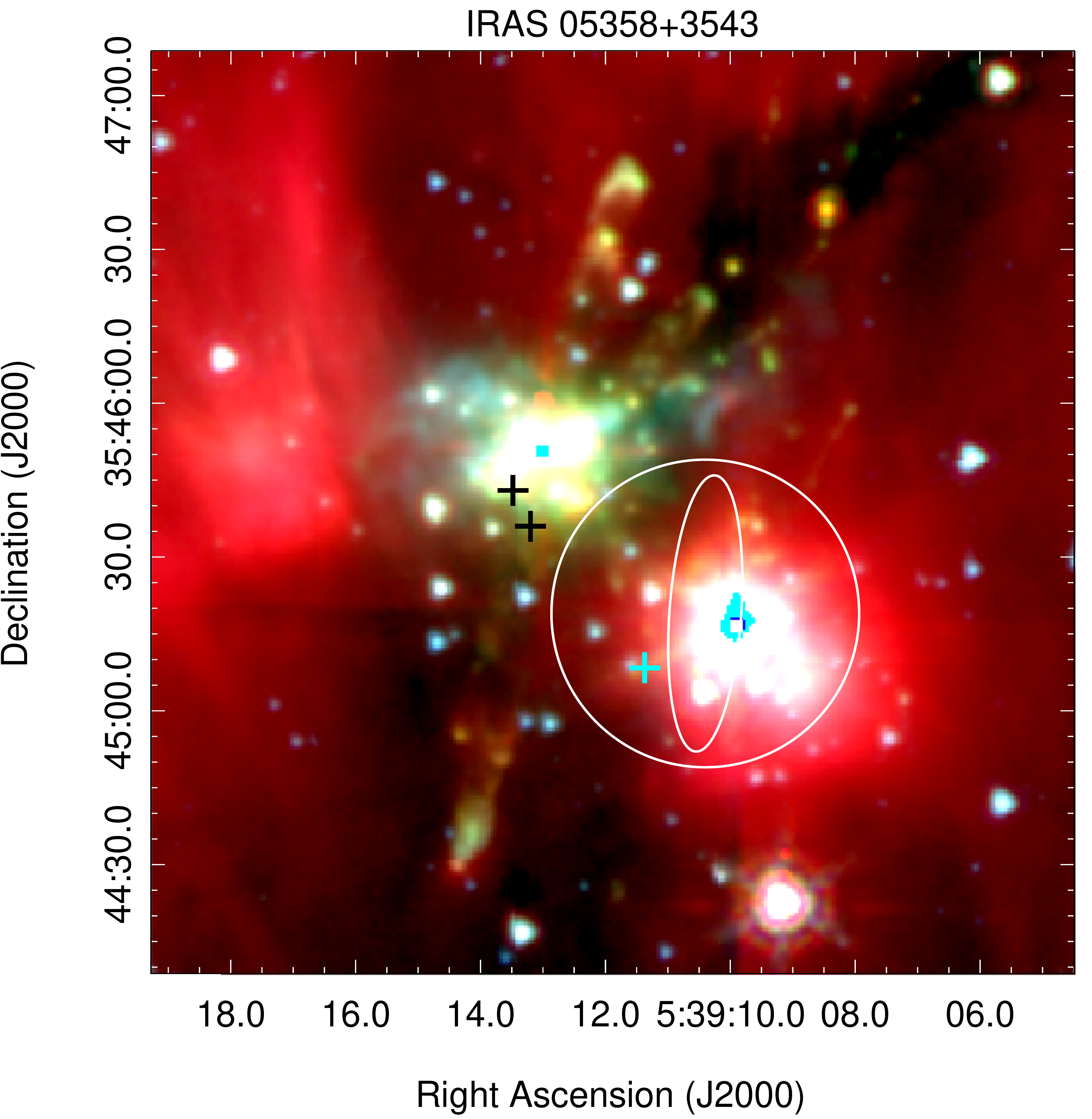} &
\includegraphics[width=70mm]{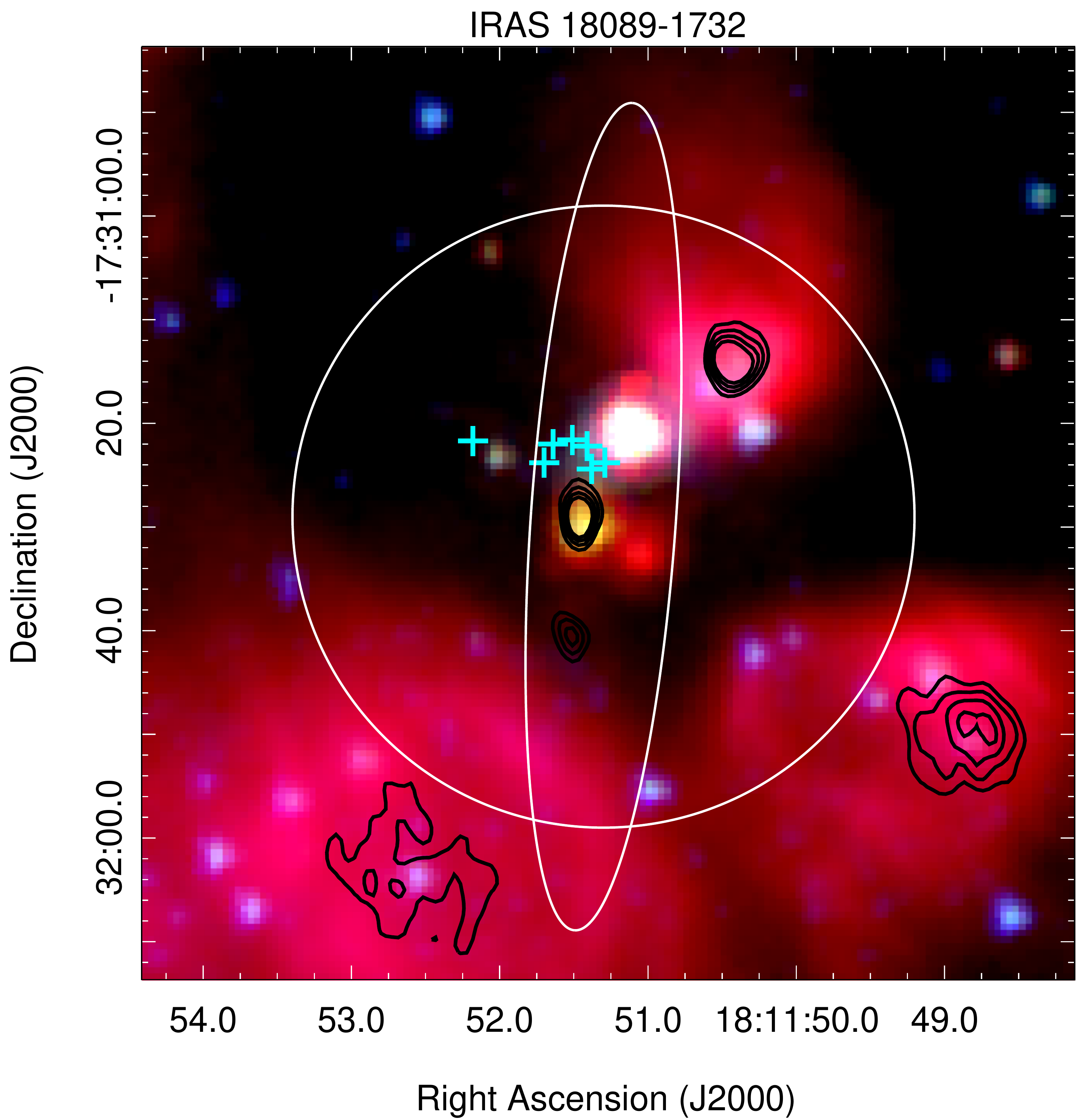} \\[0.1cm] 
\includegraphics[width=70mm]{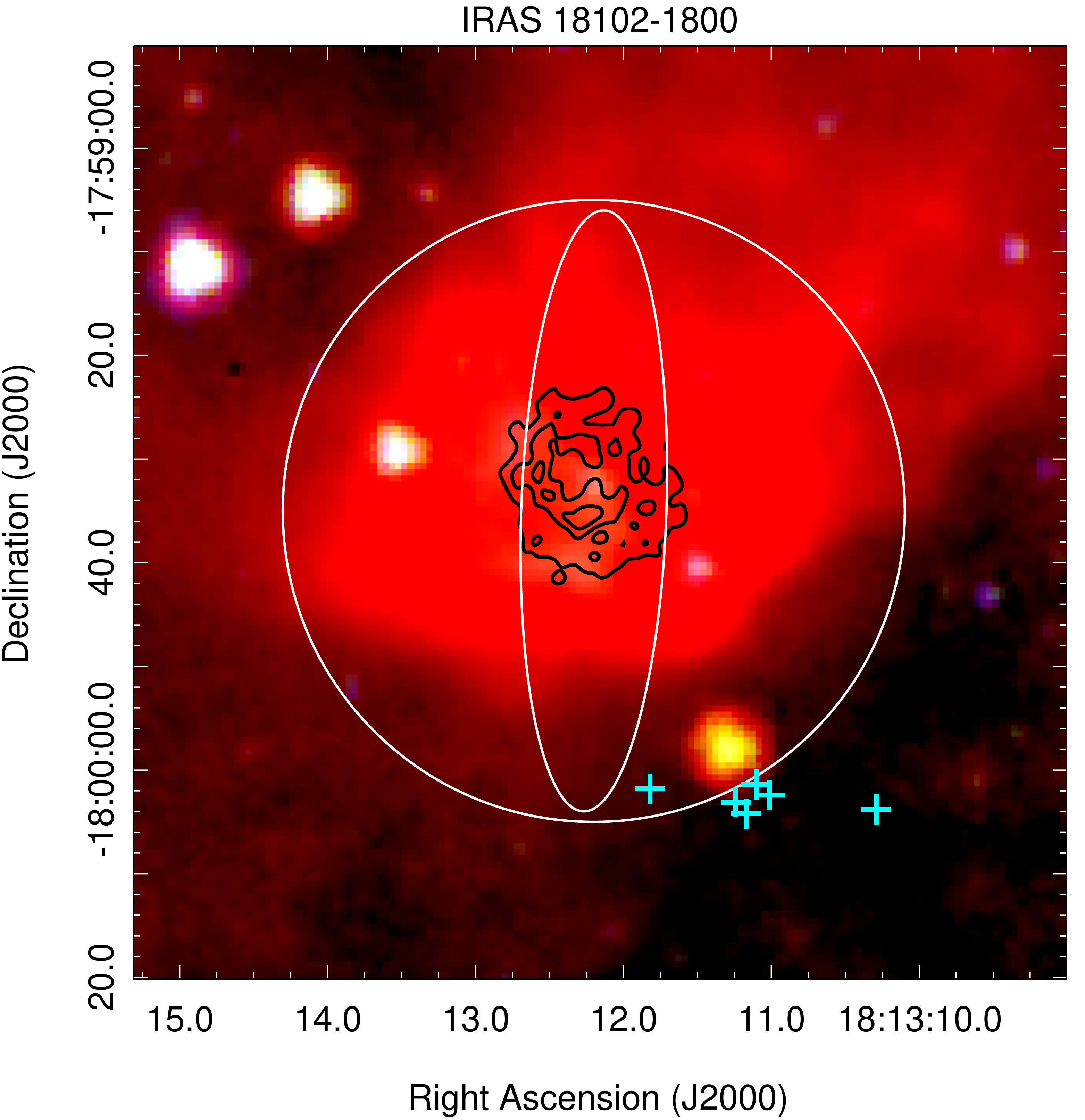} &
\includegraphics[width=70.5mm]{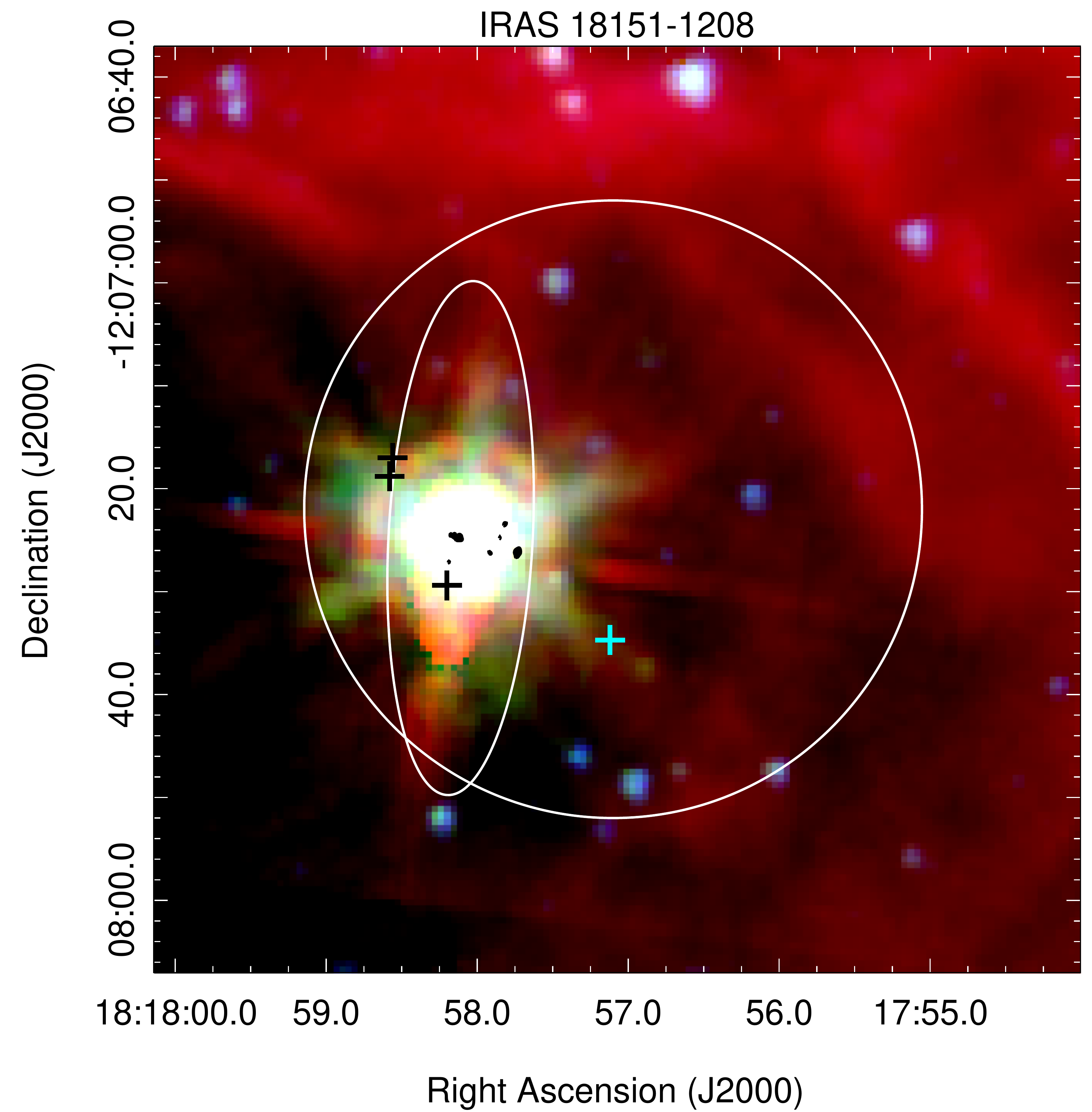} \\
\end{tabular}
\end{center}
\caption{Three-color images taken from the GLIMPSE catalog obtained with the Spitzer Space Telescope (IRAC bands: blue = 3.6 $\mu$m; green = 4.5 $\mu$m; red = 8.0 $\mu$m). Black/white contours show observations of centimeter continuum emission from VLA archival data (see Table \ref{table:refs} for contour levels and references). Cyan/black crosses mark the peak position of 44 GHz methanol masers from Table \ref{table:masers}. For sources 23033+5951 and 23139+5939 Spitzer images are not available, instead, we use mid-IR images from WISE (WISE bands: blue = 3.4 $\mu$m; green = 4.6 $\mu$m; red = 12 $\mu$m). The white circles represent the VLA primary beam of our observations ($\sim 1'$). The white ellipses are centered at the IRAS position which in some cases is offset by several arcseconds from the positions reported by S02.}
\label{irac}
\end{figure*}

\begin{figure*}[h!]
\addtocounter{figure}{-1}
\begin{center}
\begin{tabular}{ll}
\includegraphics[width=70mm]{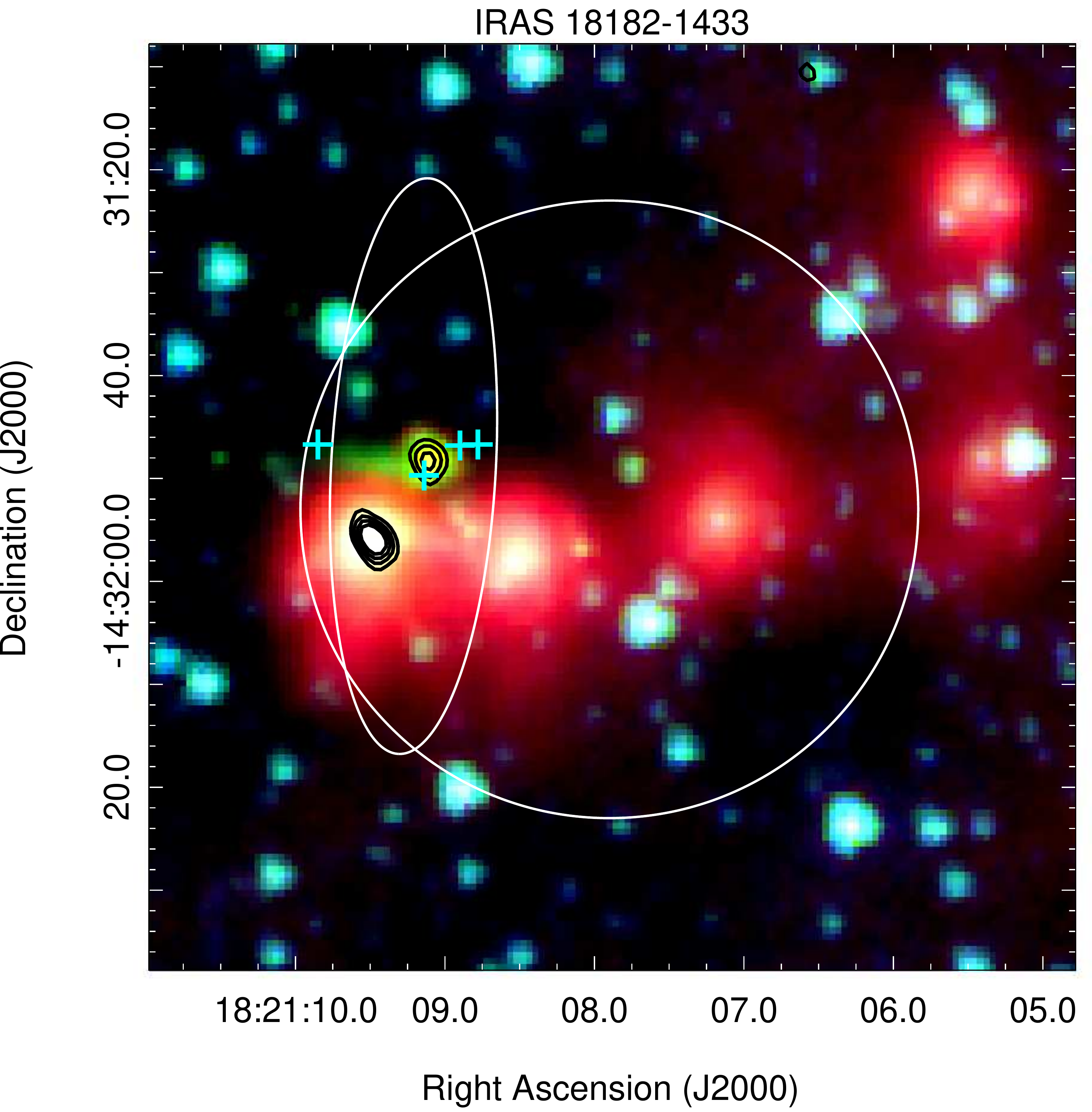} &
\includegraphics[width=70mm]{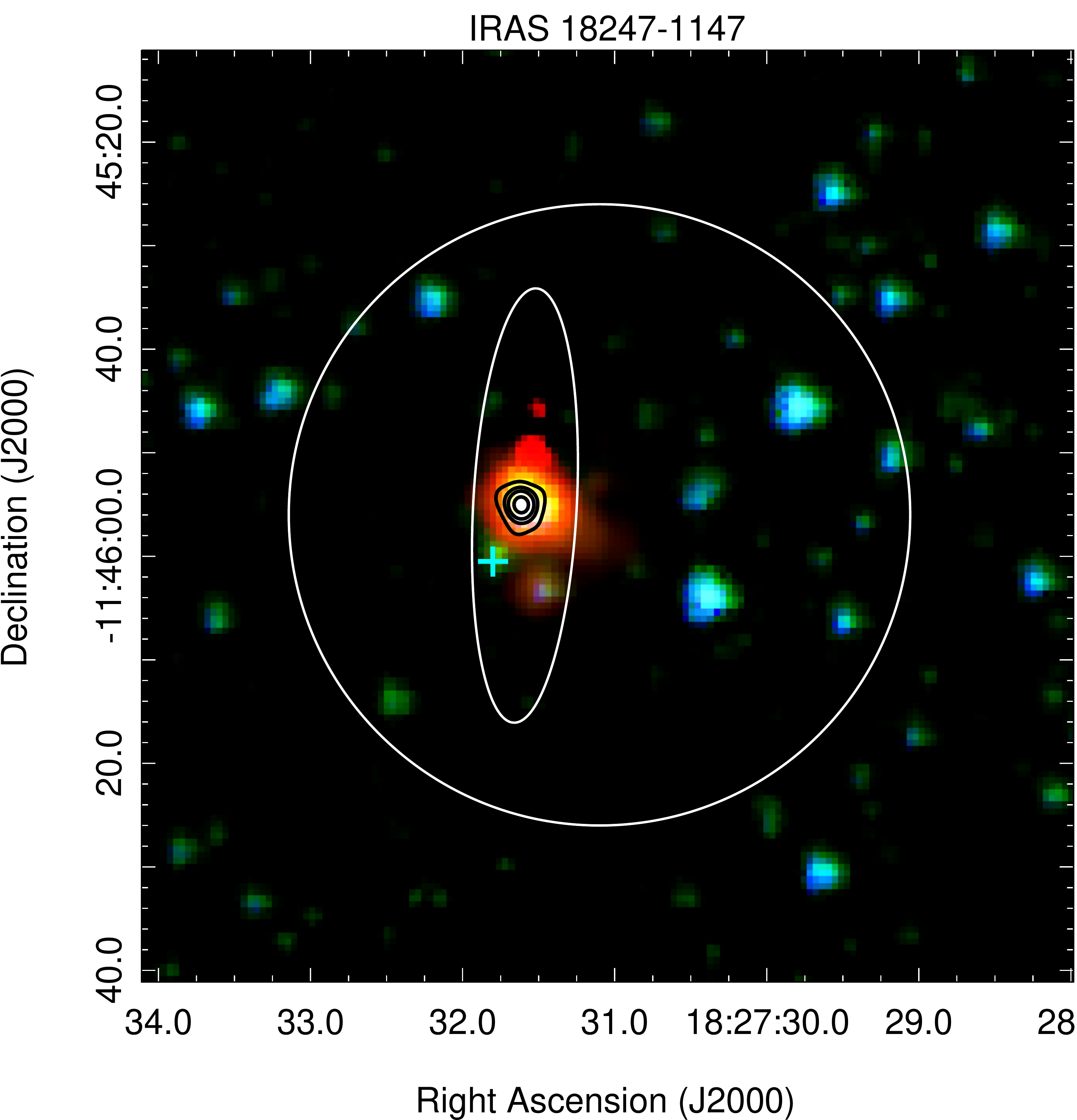} \\[0.1cm] 
\includegraphics[width=70mm]{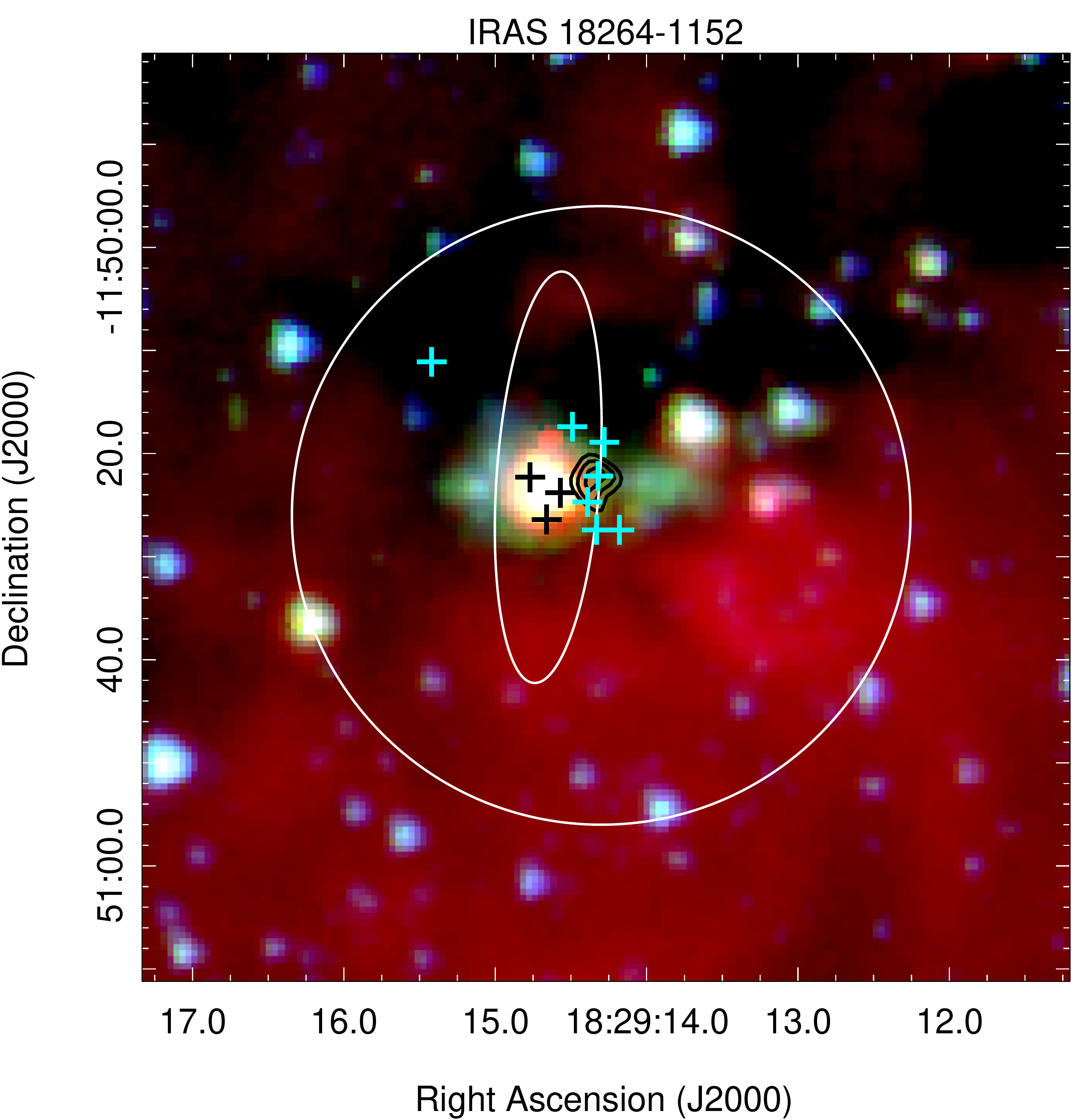} &
\includegraphics[width=70mm]{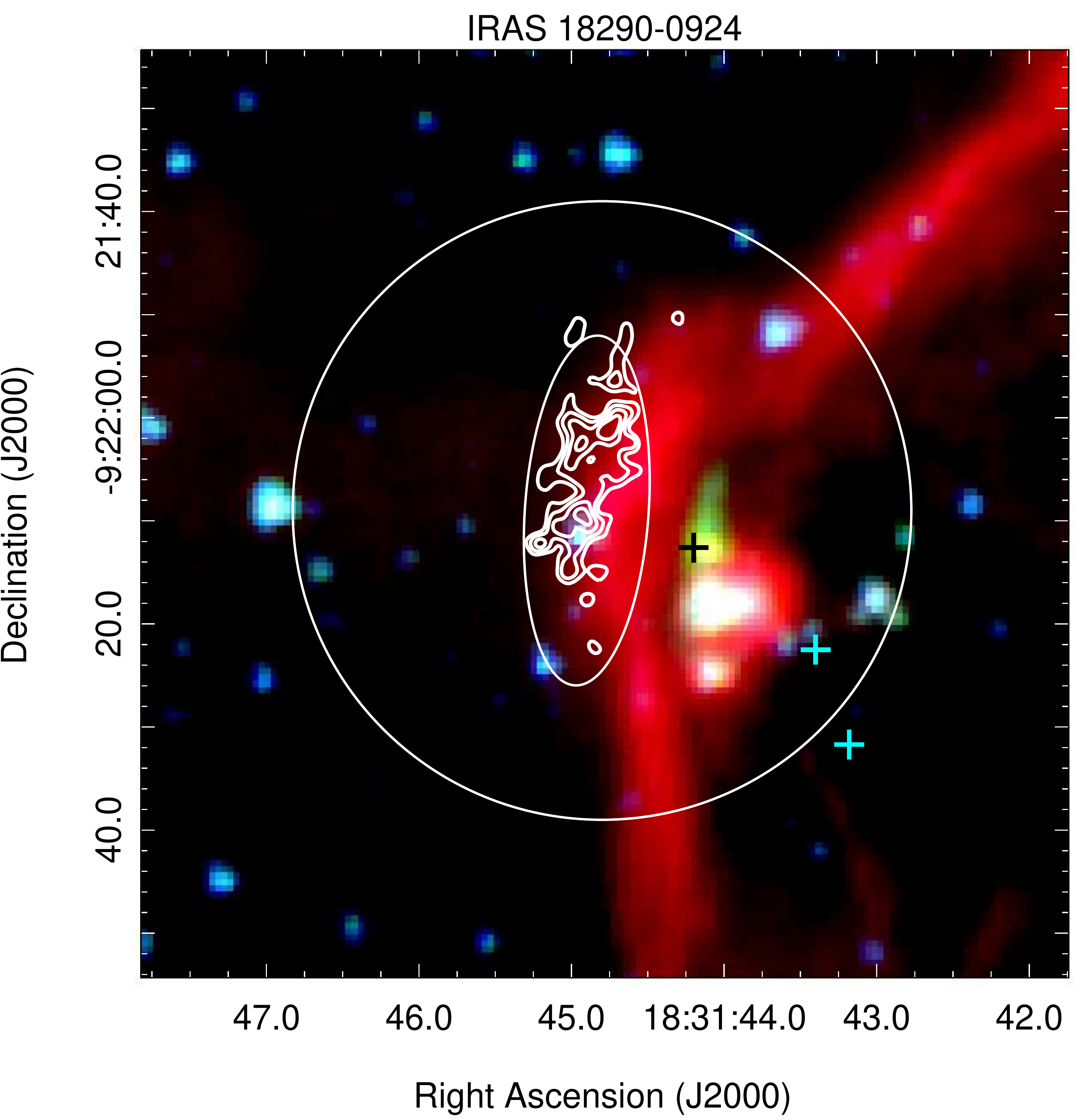} \\[0.1cm]
\includegraphics[width=70mm]{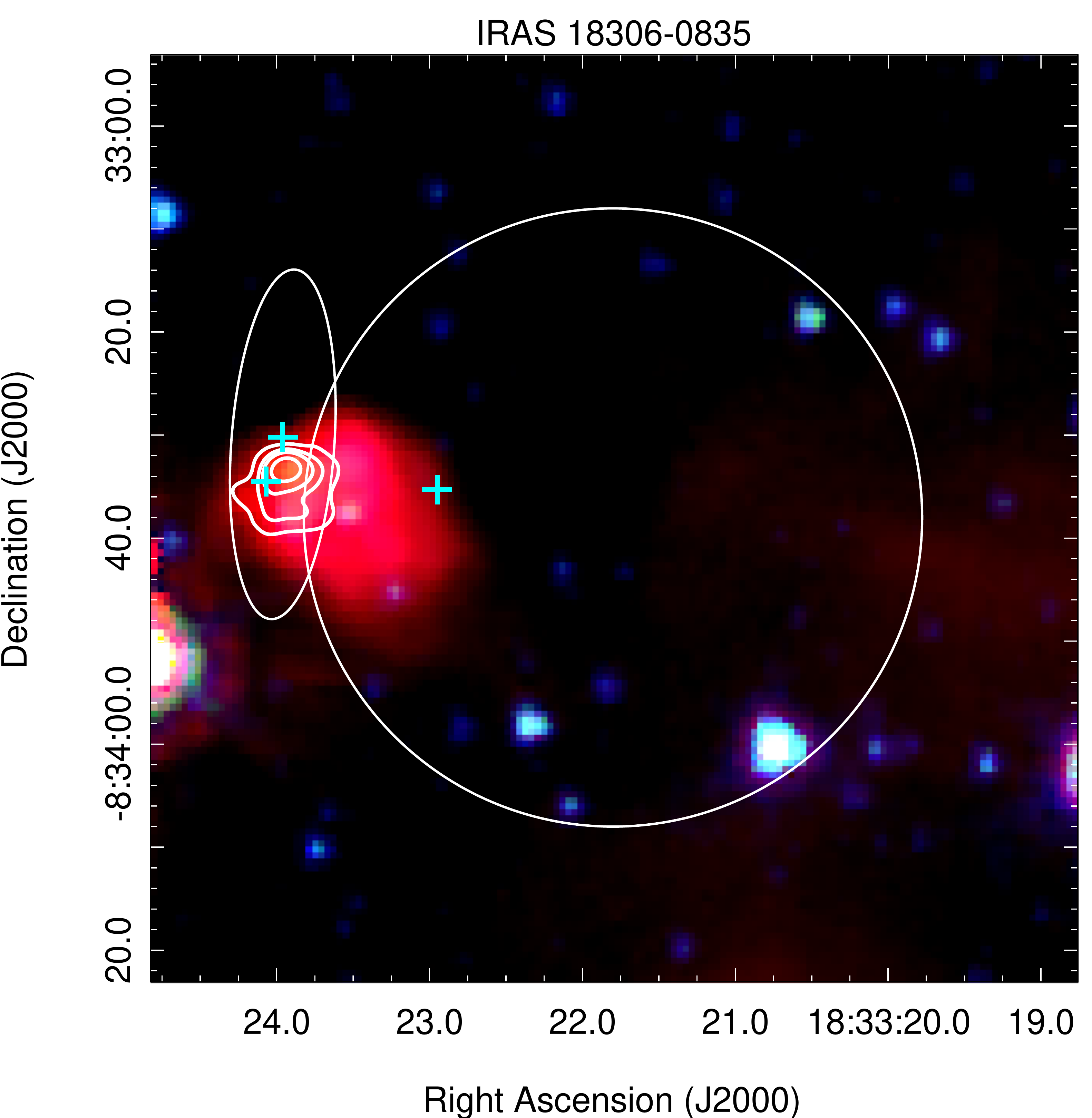} &
\includegraphics[width=70mm]{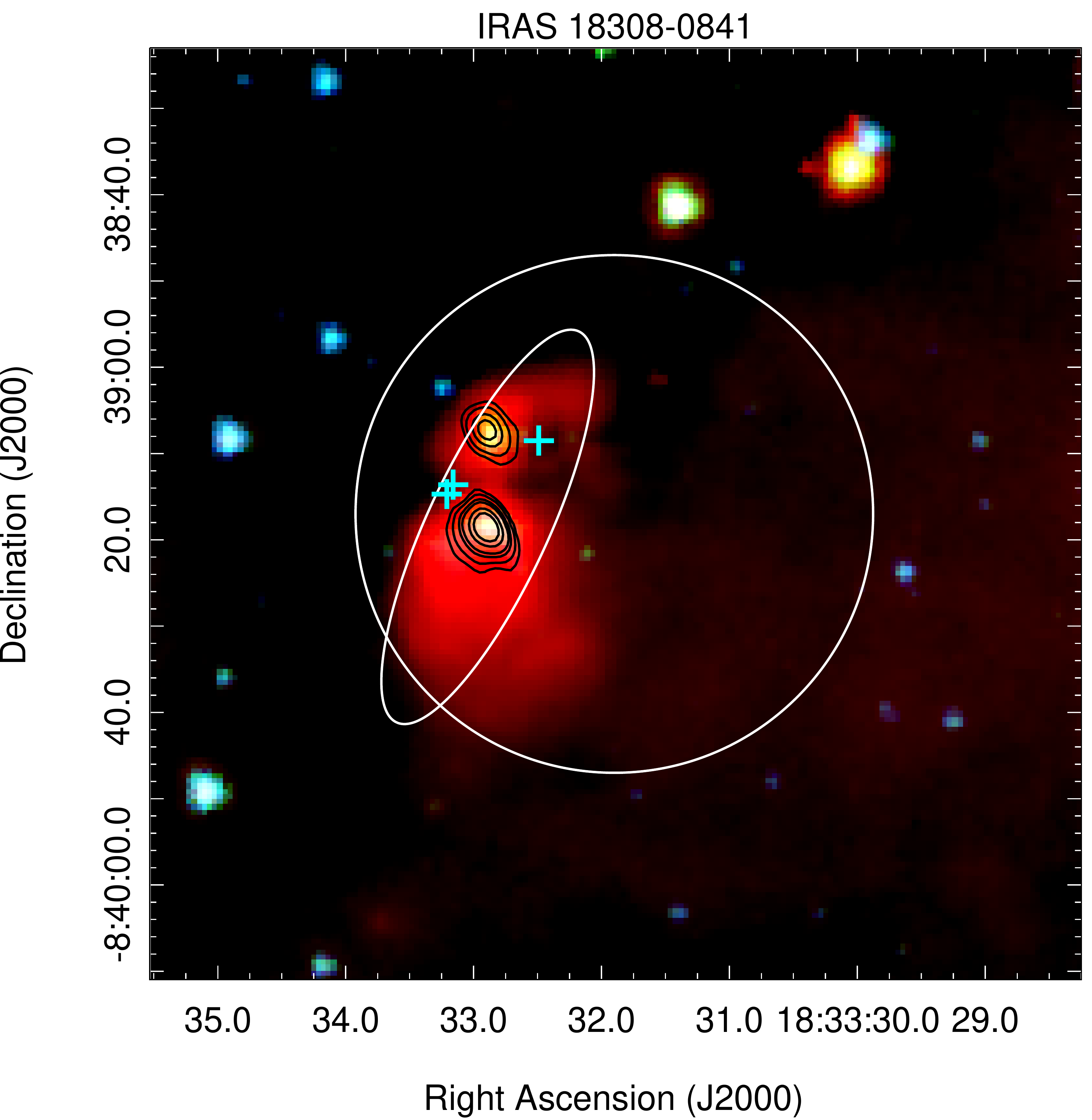} 
\end{tabular}
\end{center}
\caption{Continued}
\label{irac}
\end{figure*}

\begin{figure*}[h!]
\addtocounter{figure}{-1}
\begin{center}
\begin{tabular}{ll}
\includegraphics[width=70mm]{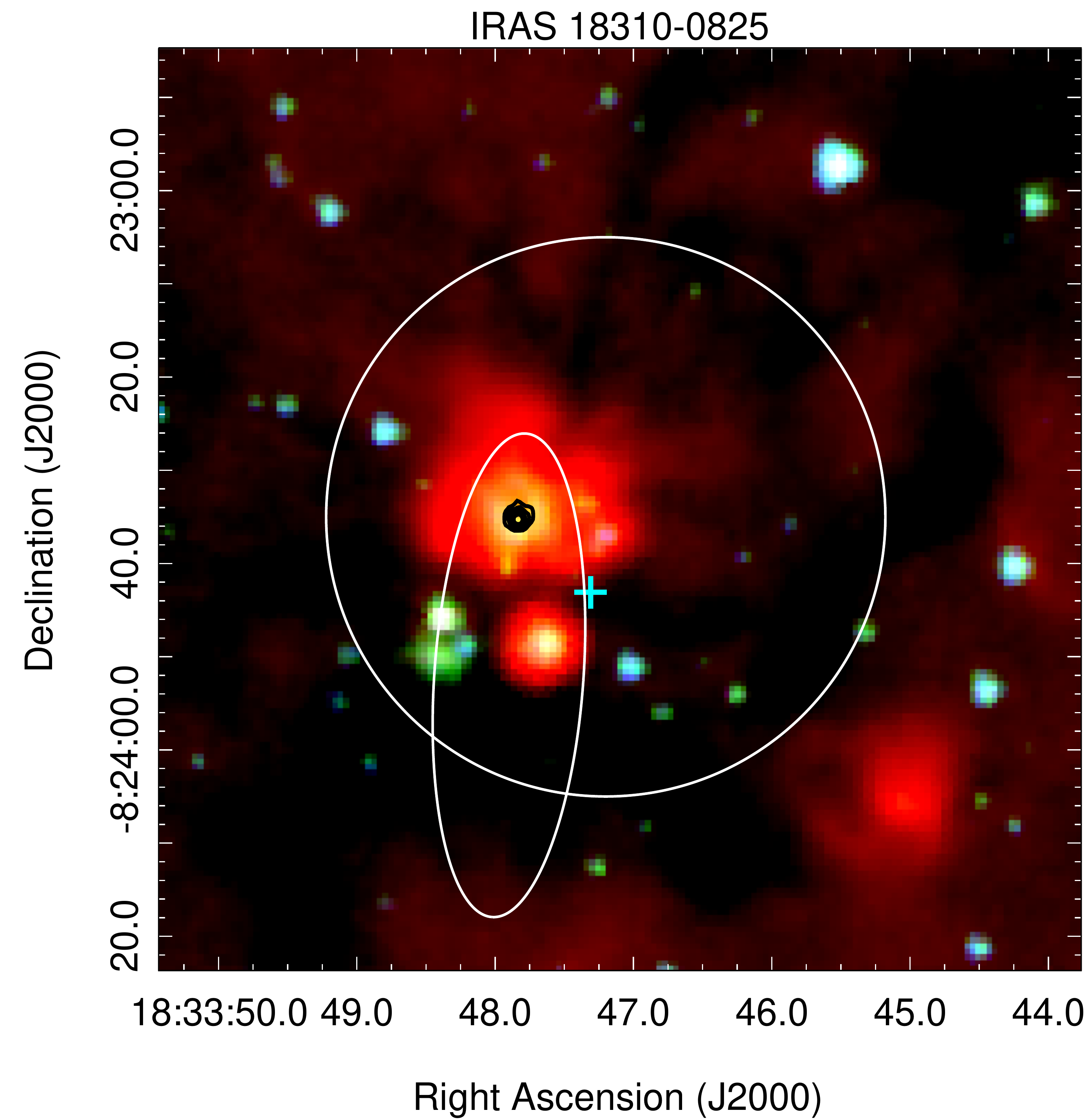} &
\includegraphics[width=70mm]{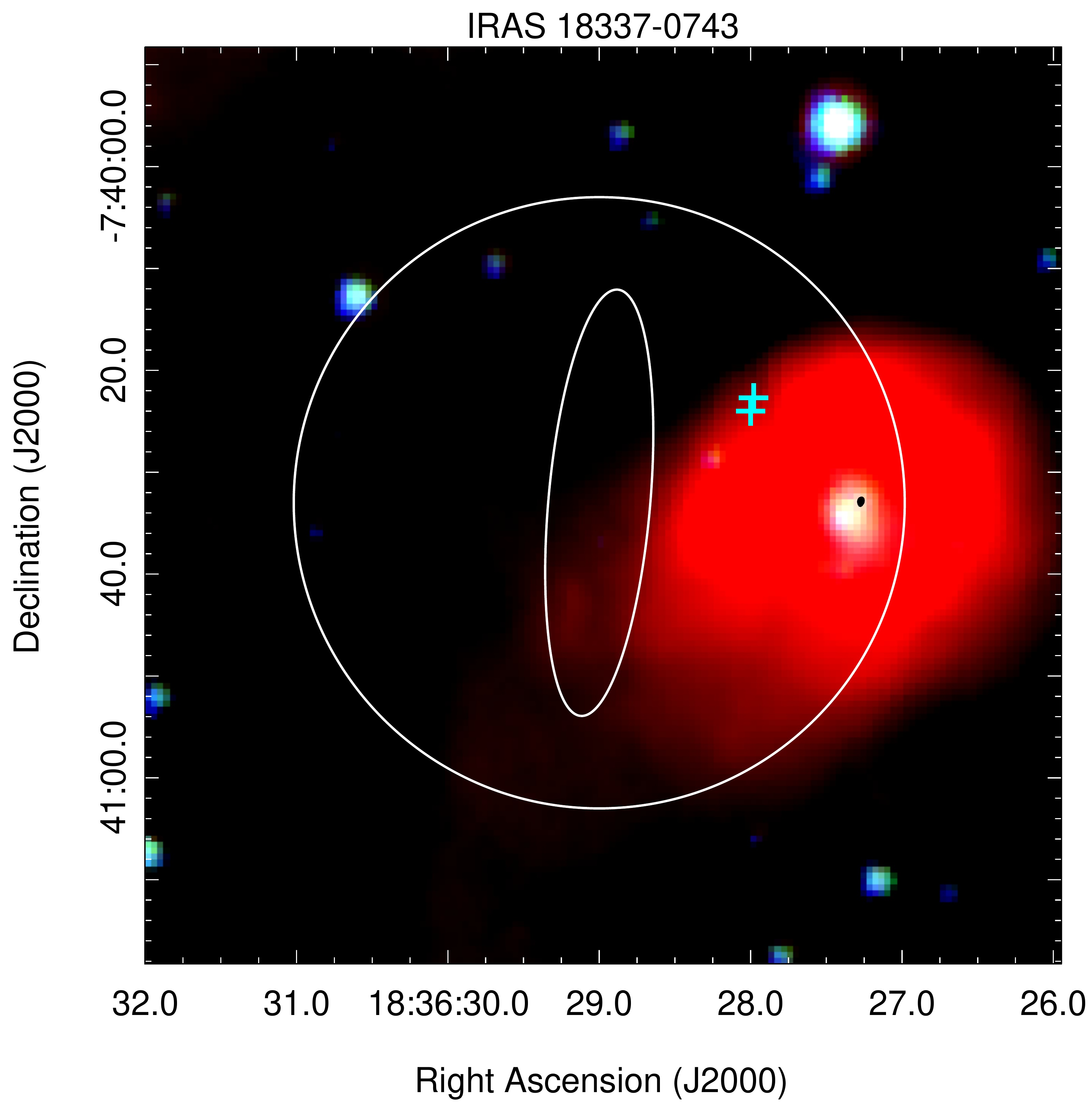} \\[0.1cm]
\includegraphics[width=70mm]{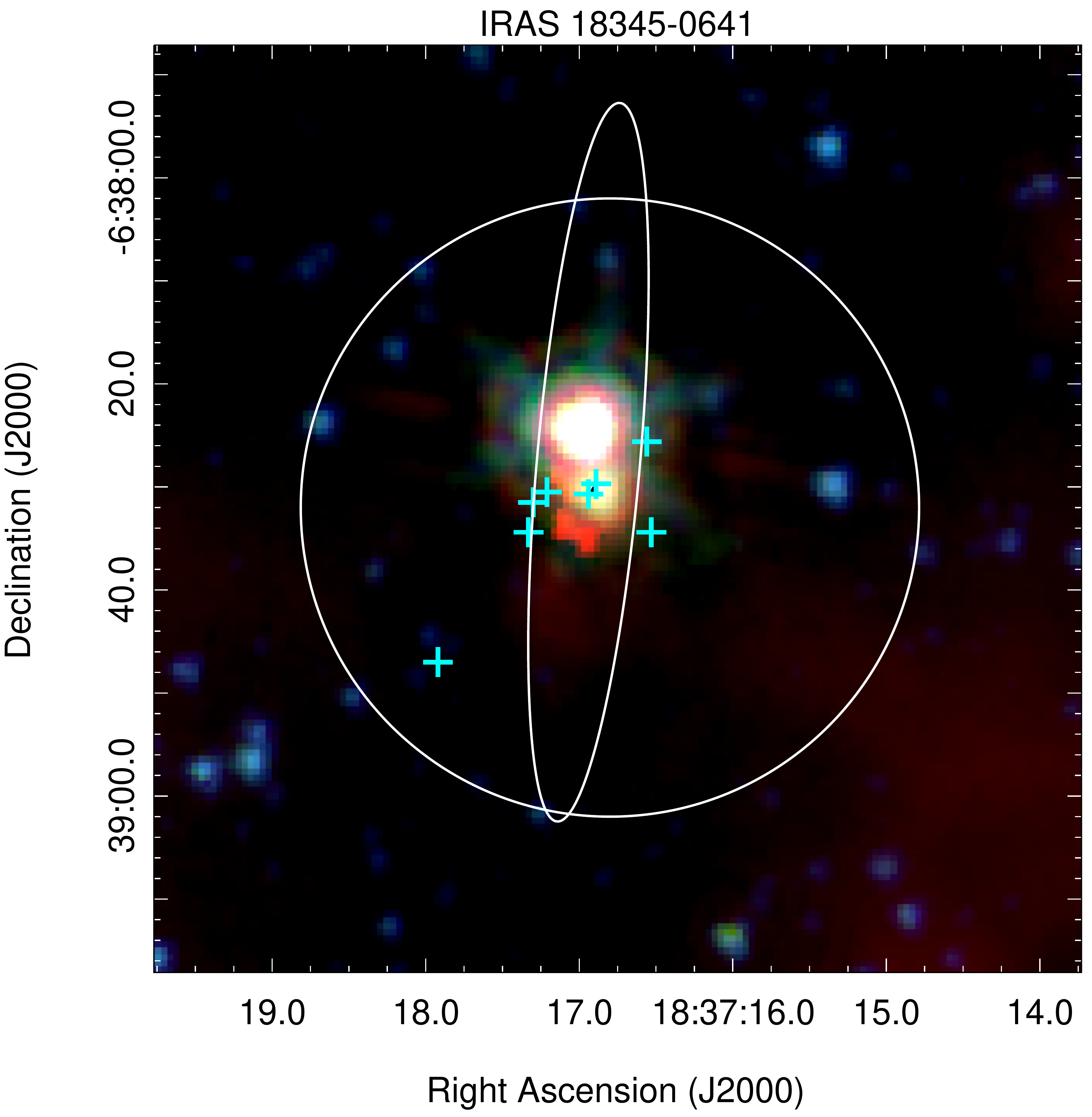} &
\includegraphics[width=70mm]{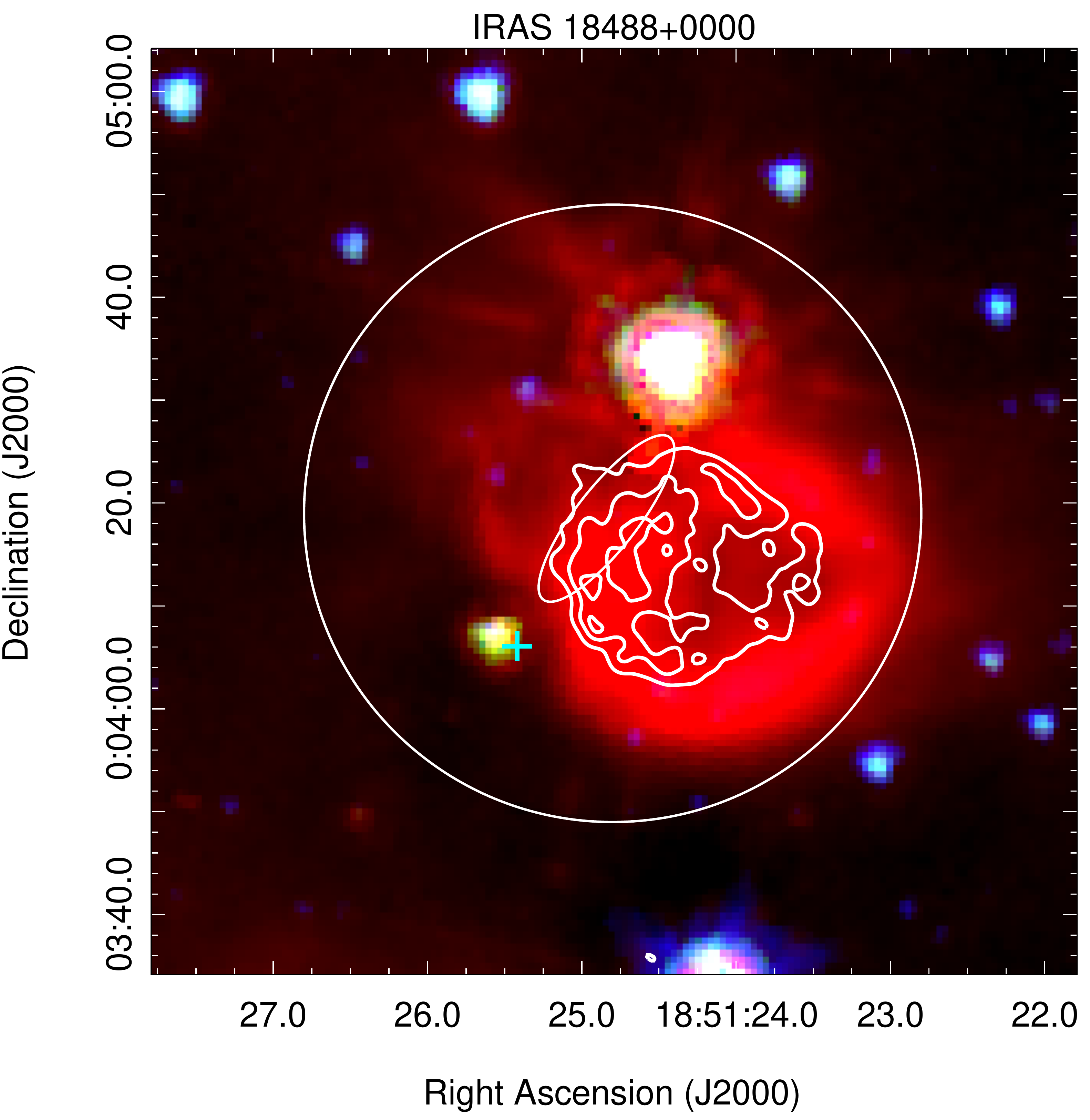} \\[0.1cm]
\includegraphics[width=70mm]{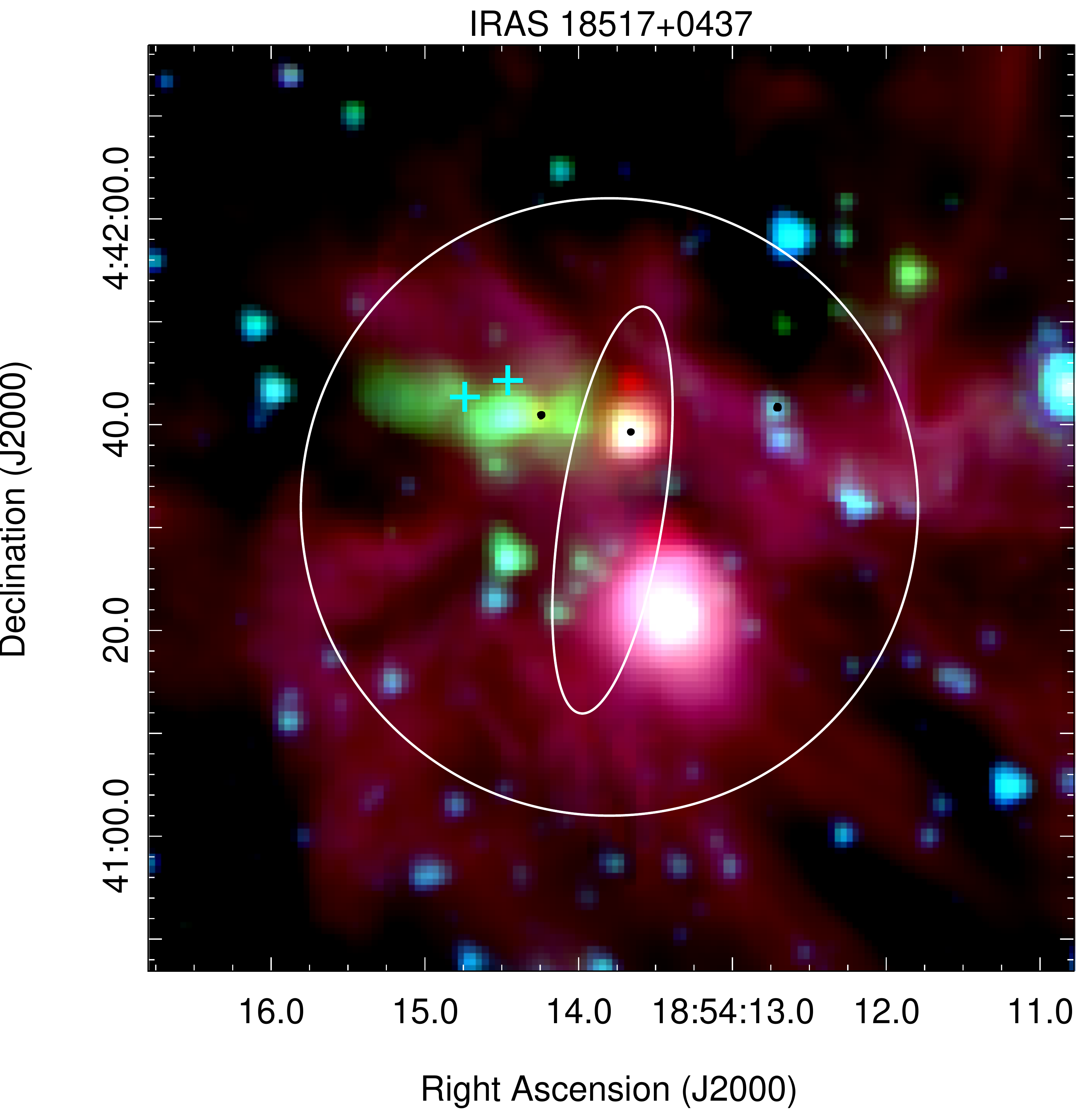} &
\includegraphics[width=70mm]{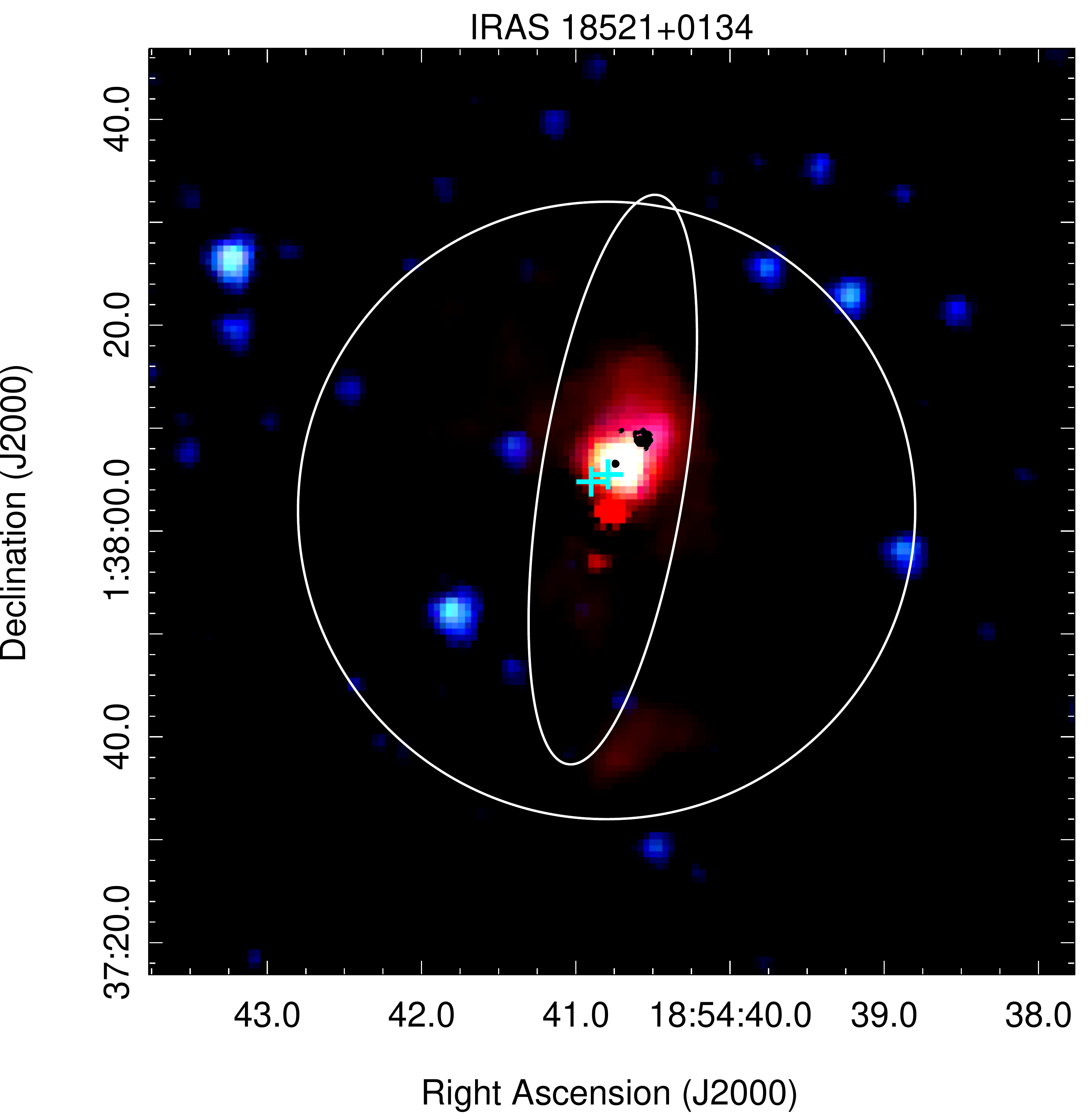} 
\end{tabular}
\end{center}
\caption{Continued}
\label{irac}
\end{figure*}

\begin{figure*}[h!]
\addtocounter{figure}{-1}
\begin{center}
\begin{tabular}{cc}
\includegraphics[width=70mm]{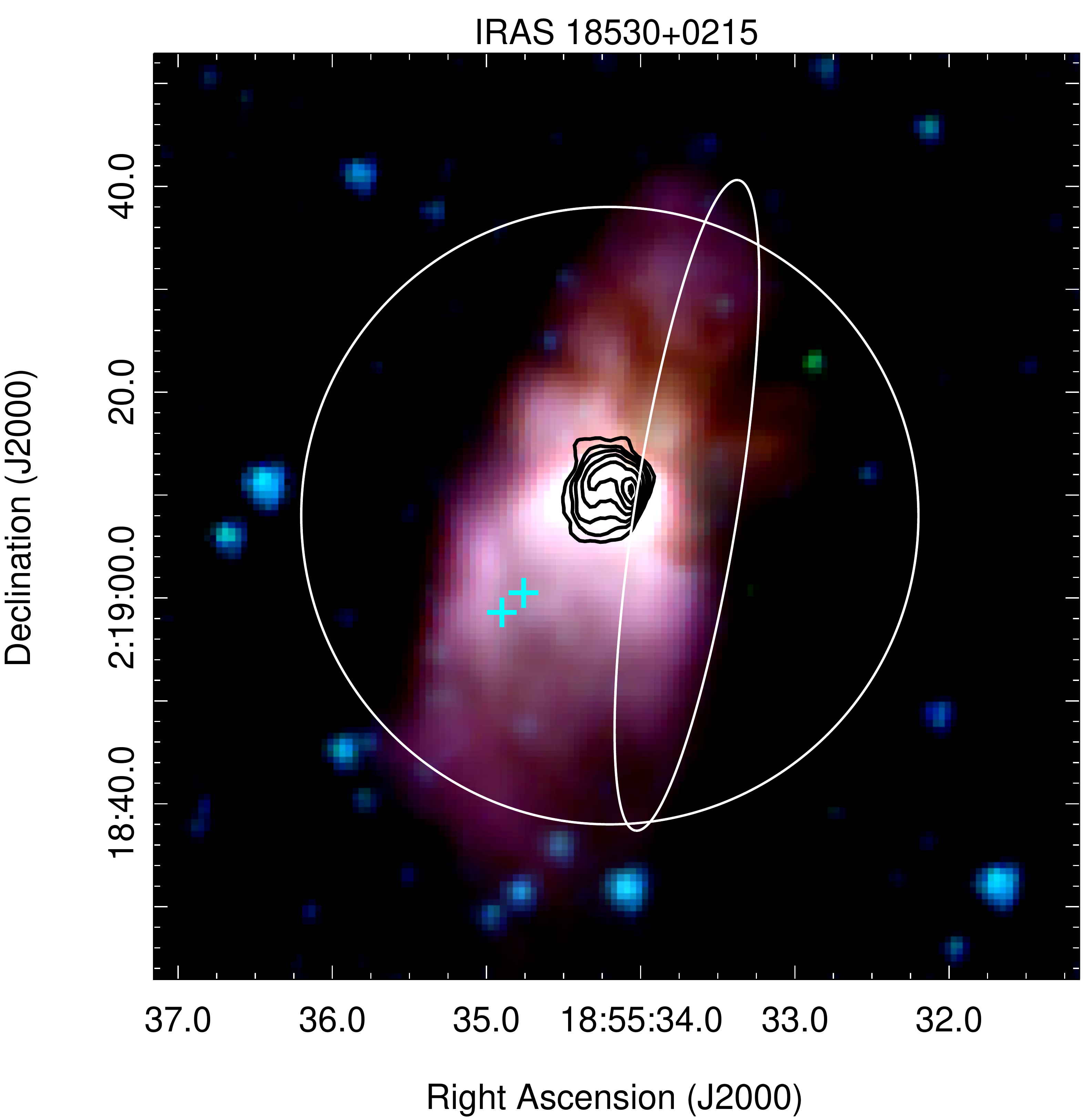} &
\includegraphics[width=70mm]{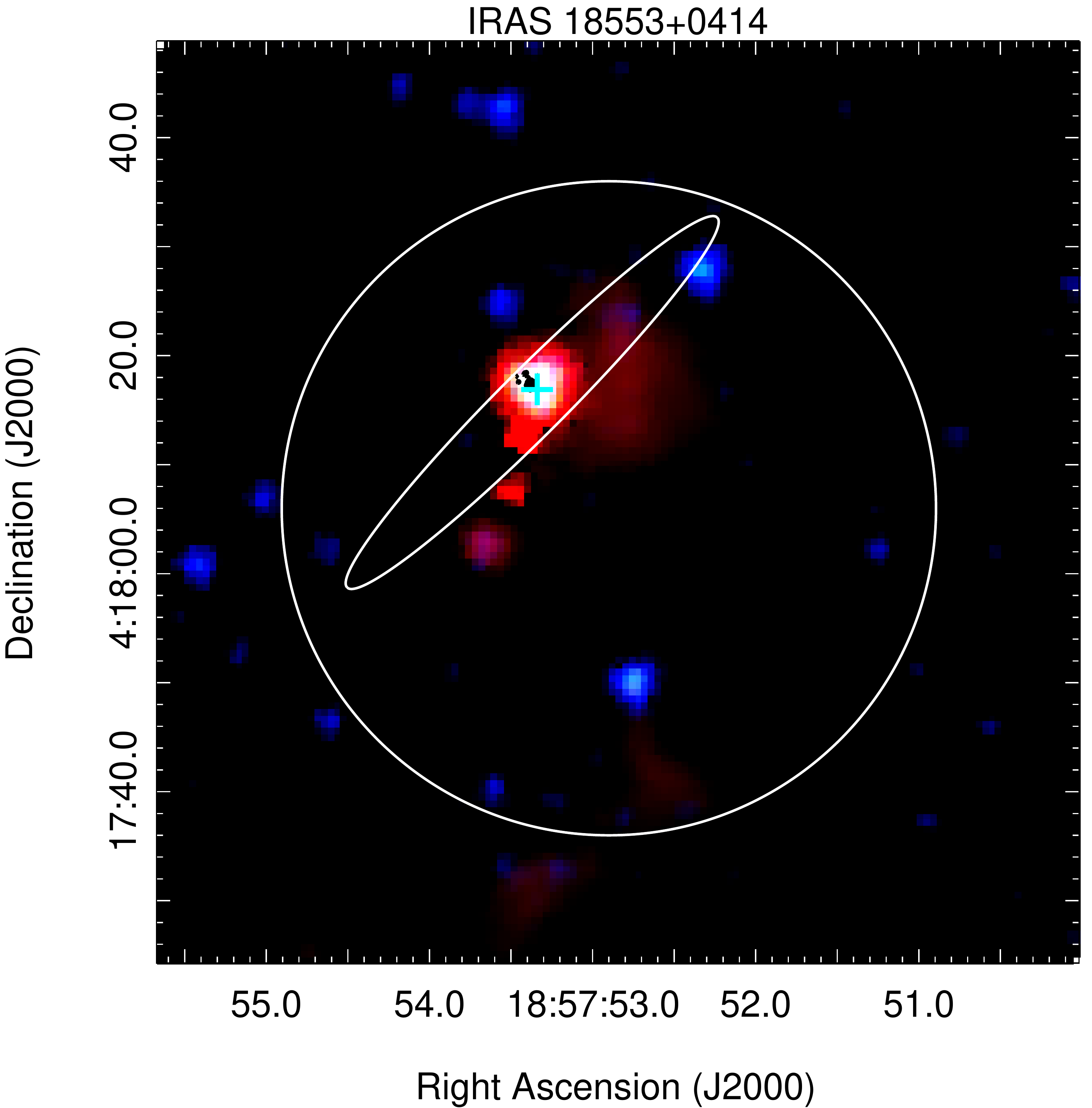} \\[0.1cm]
\includegraphics[width=70mm]{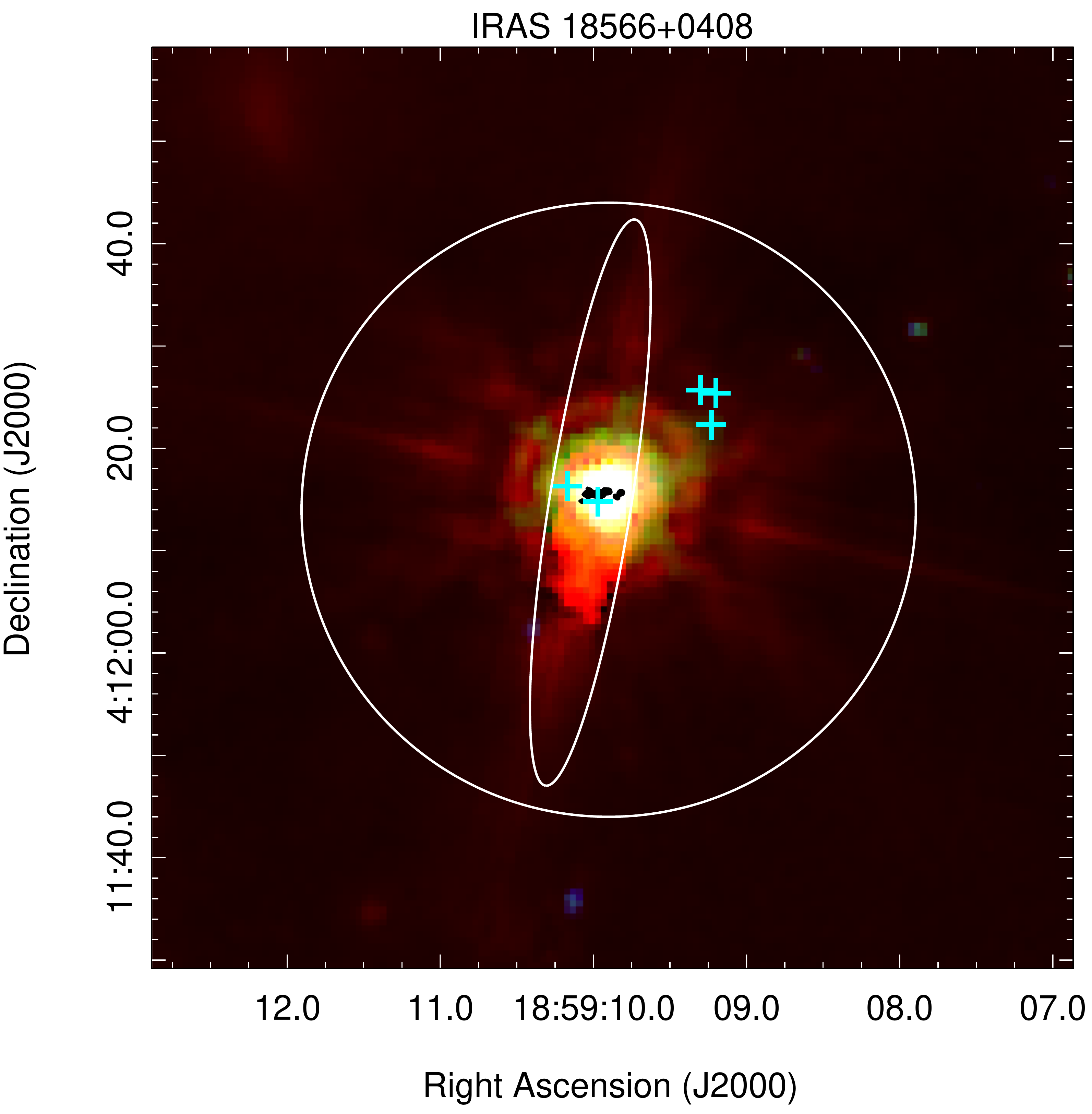} &
\includegraphics[width=70mm]{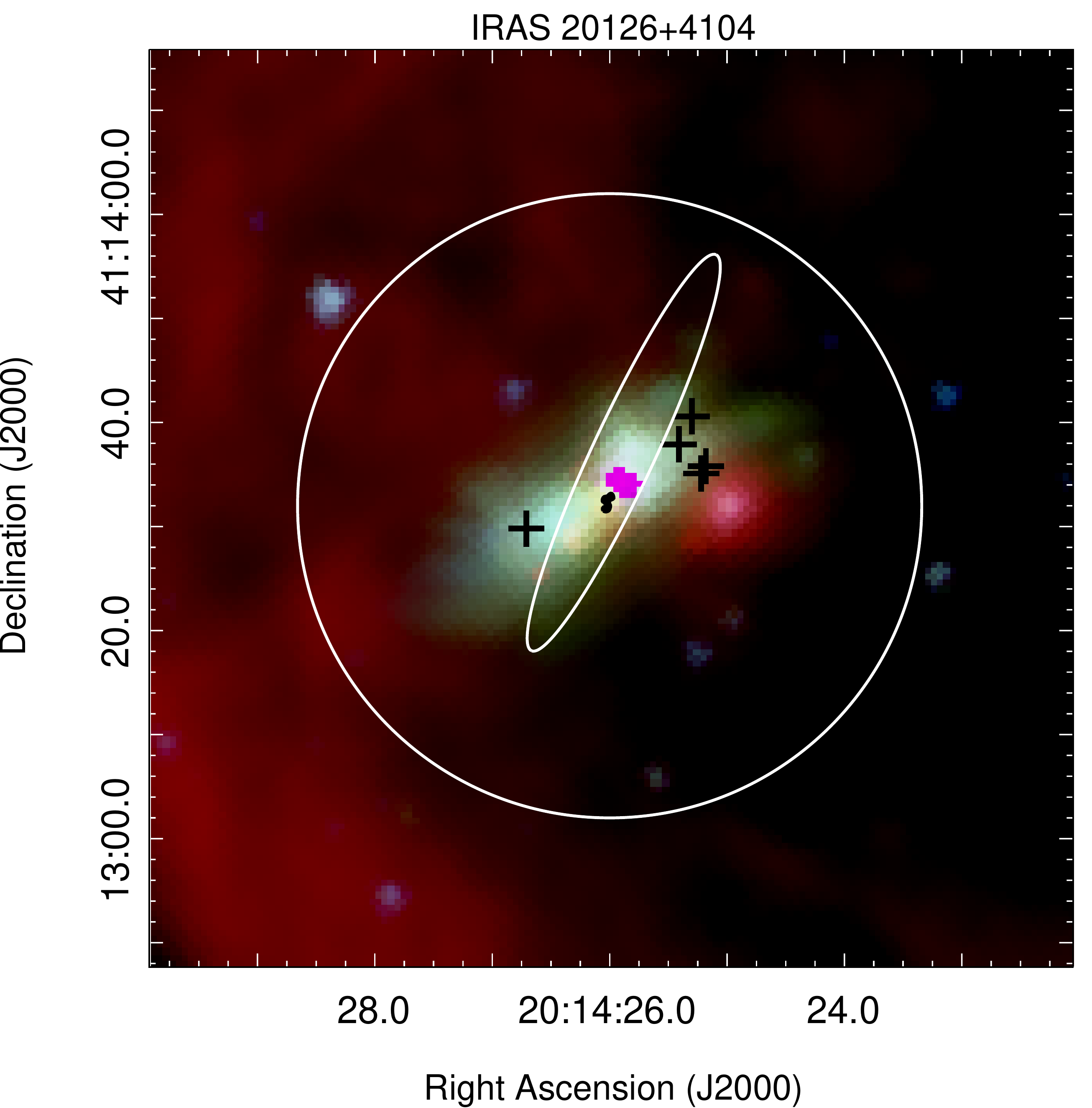} \\[0.1cm]
\includegraphics[width=70mm]{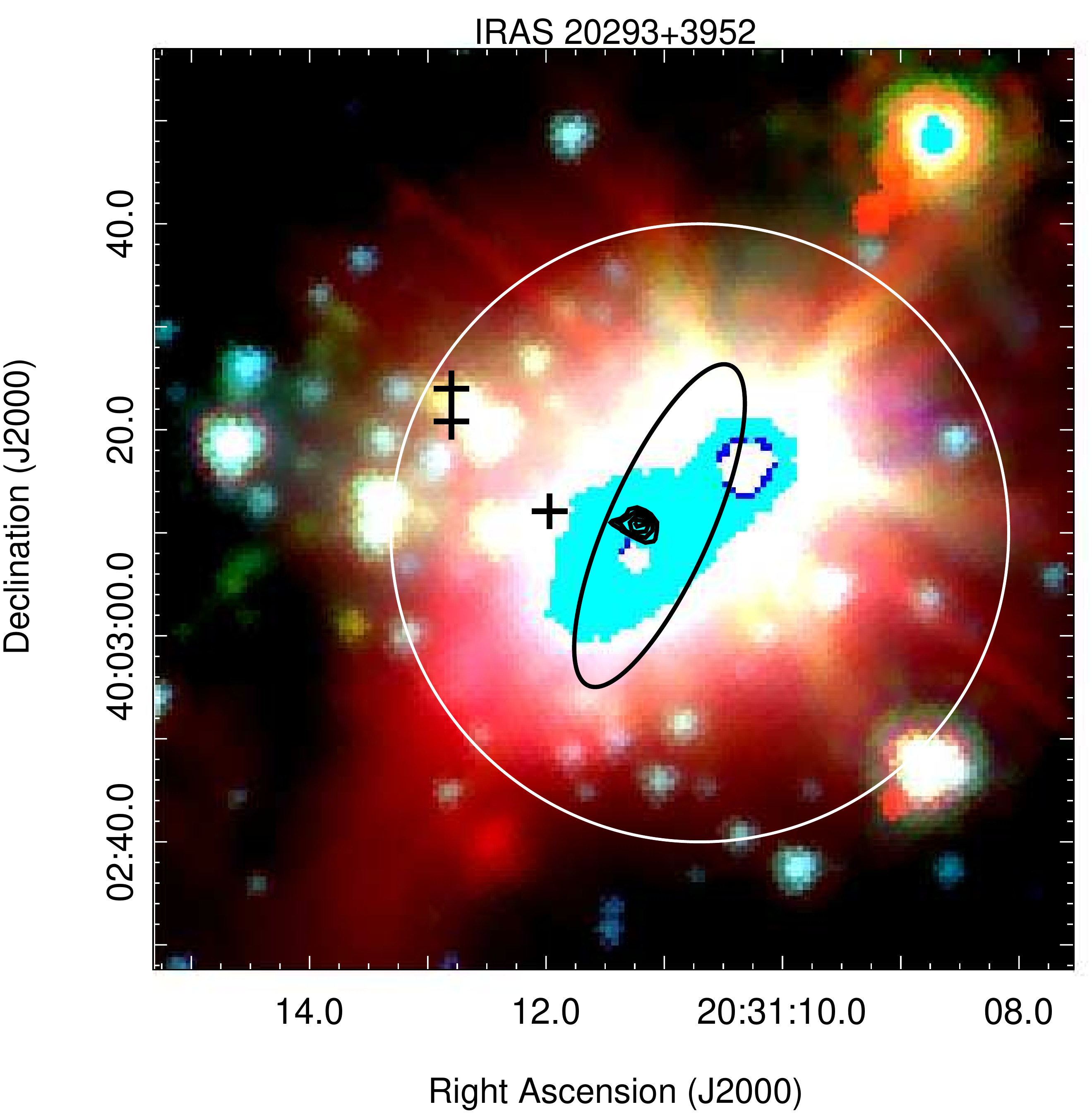} &
\includegraphics[width=70mm]{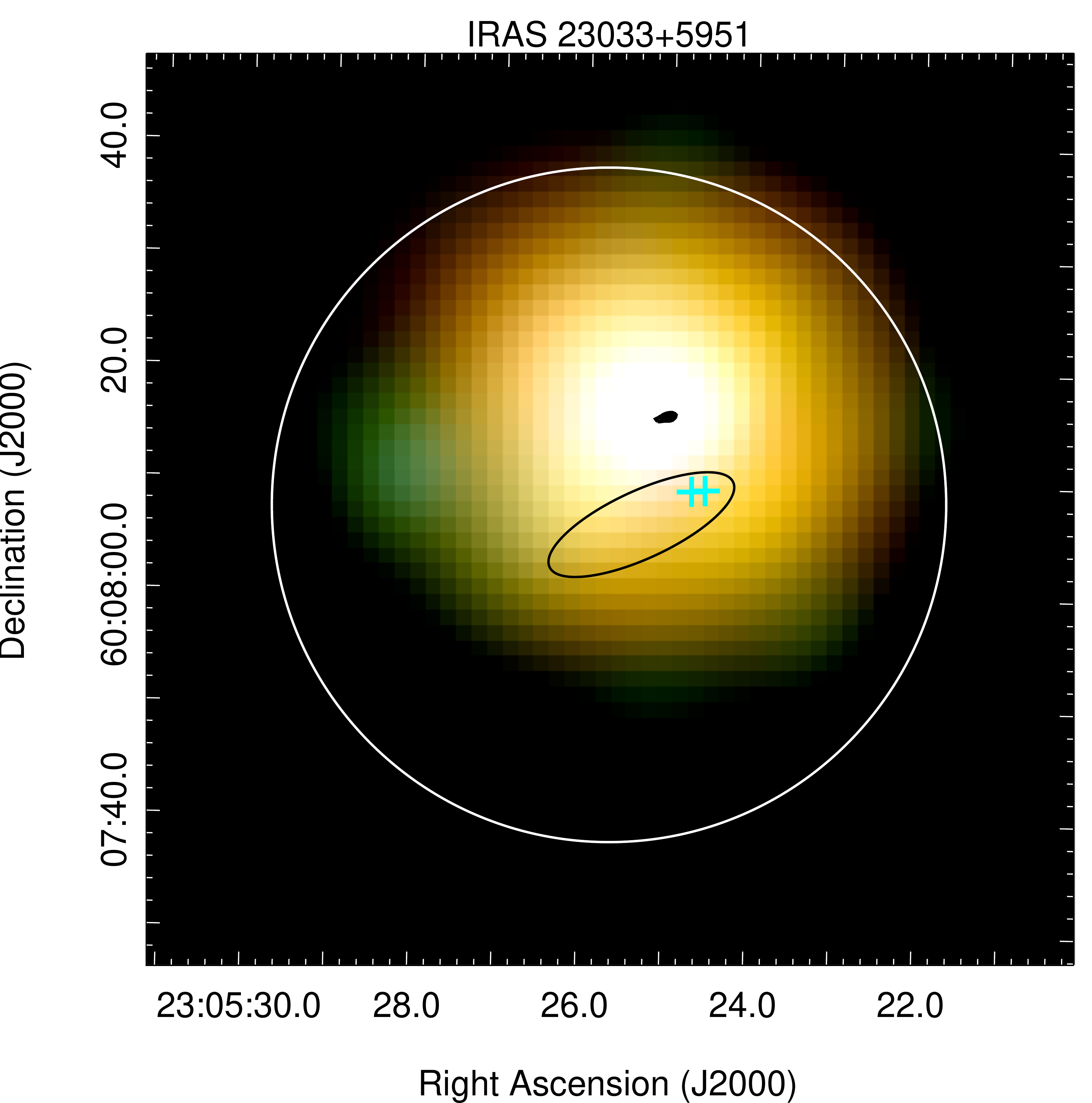} 
\end{tabular}
\caption{Continued}
\label{irac}
\end{center}
\end{figure*}

\begin{figure*}[h!]
\addtocounter{figure}{-1}
\begin{center}
\begin{tabular}{cc}
\includegraphics[width=80mm]{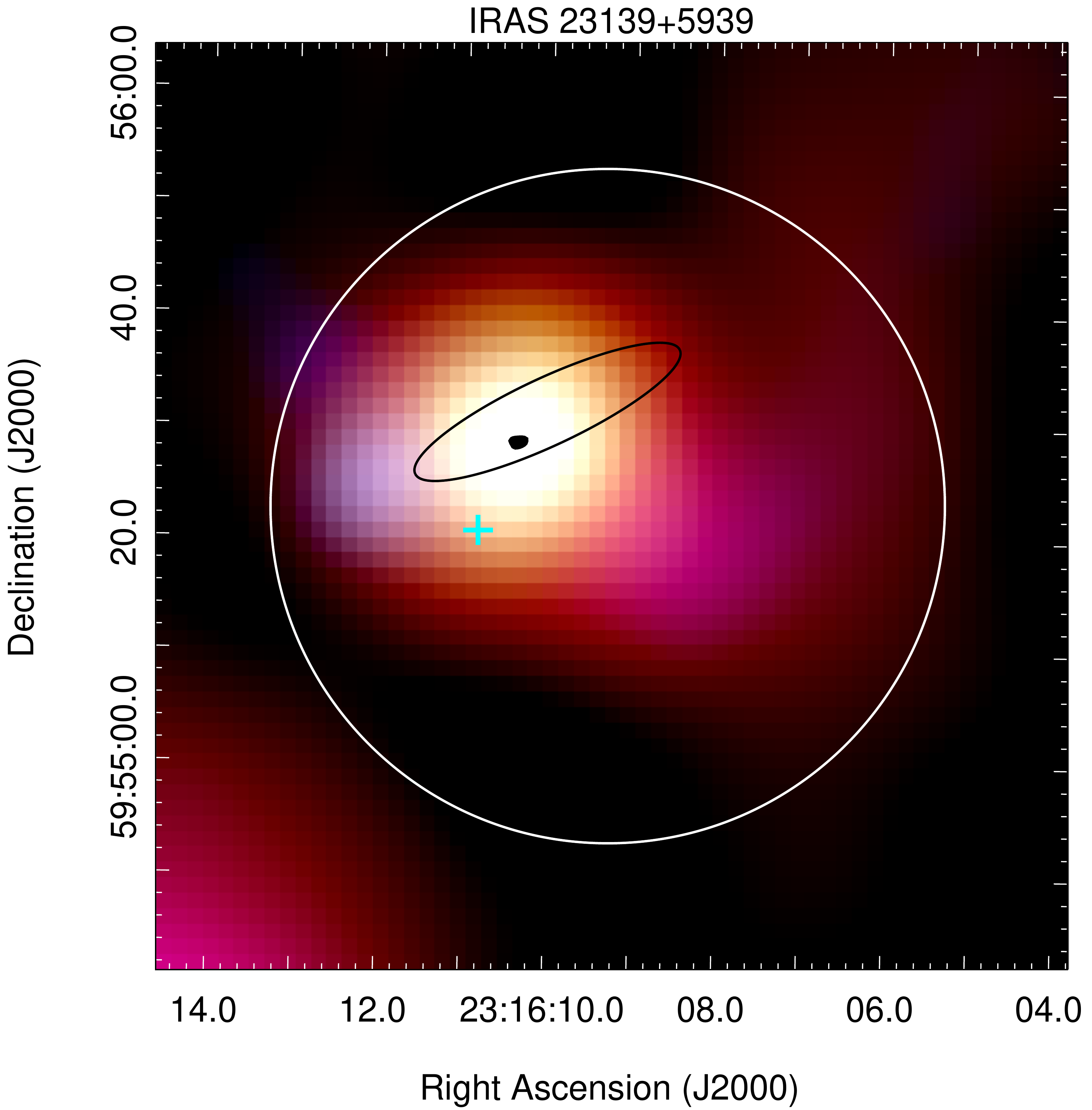} &
\includegraphics[width=80mm]{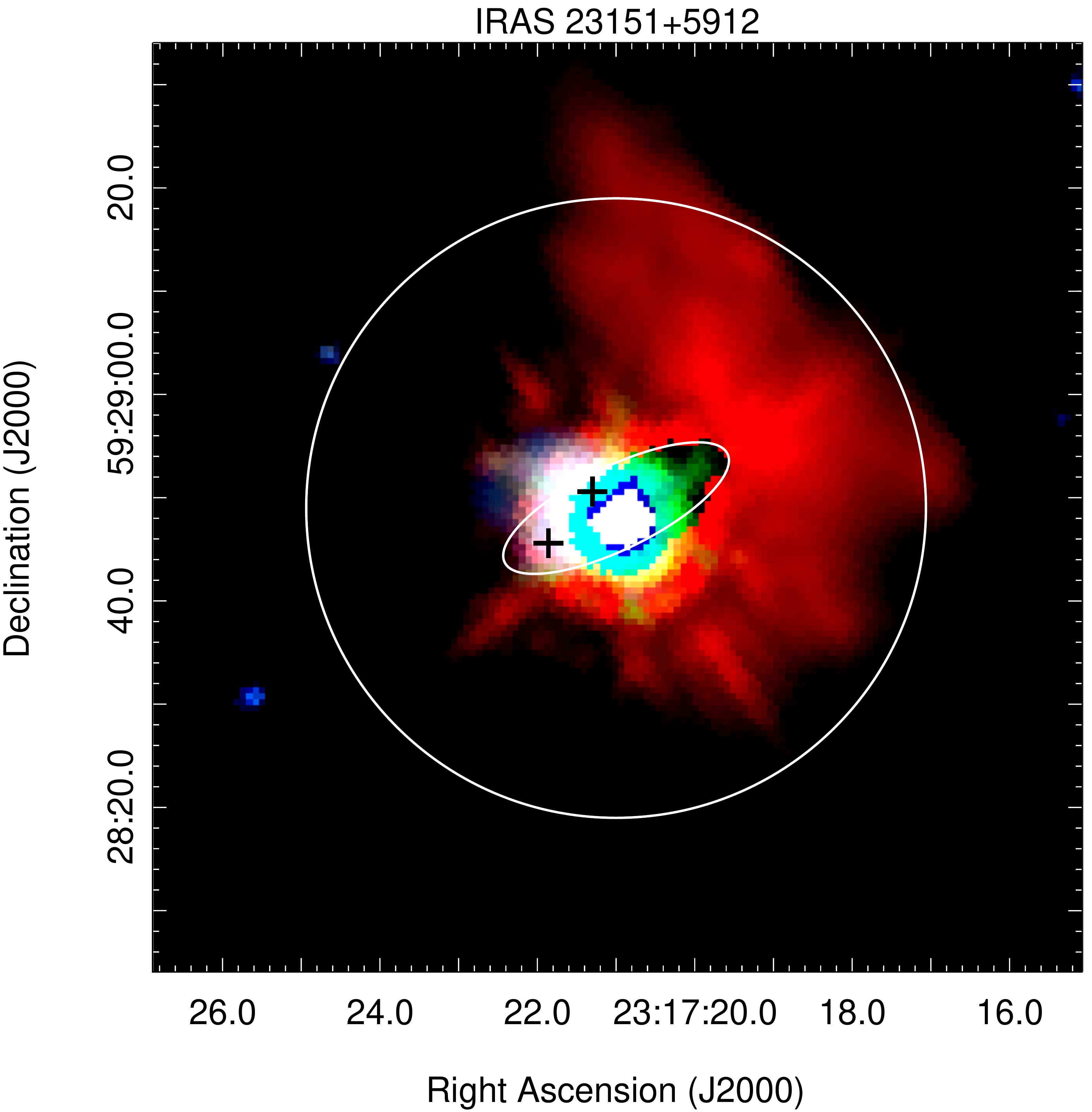} 
\end{tabular}
\caption{Continued}
\label{irac}
\end{center}
\end{figure*}

\section{Results \label{sec:res}}

\begin{figure}
\plottwo{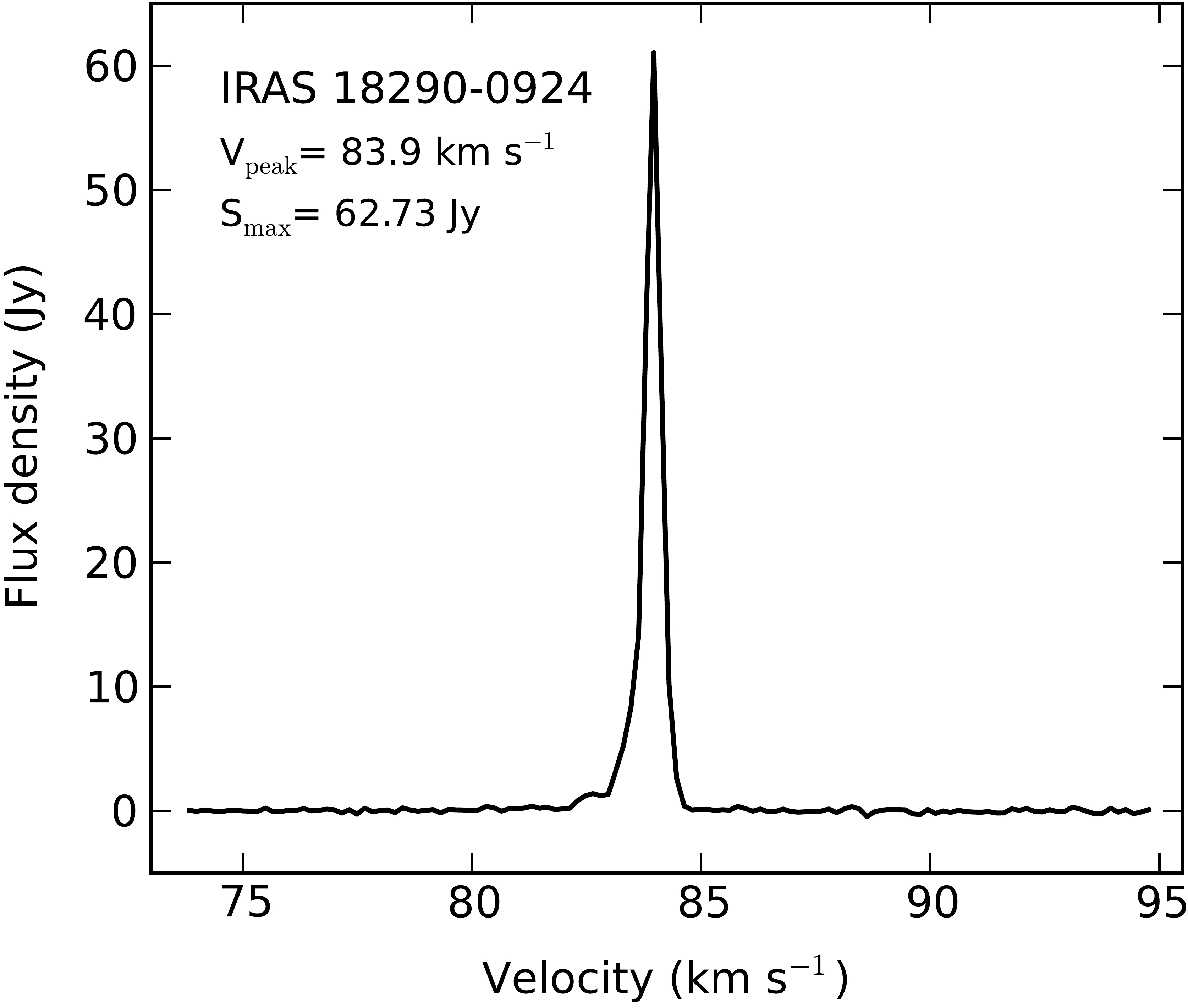}{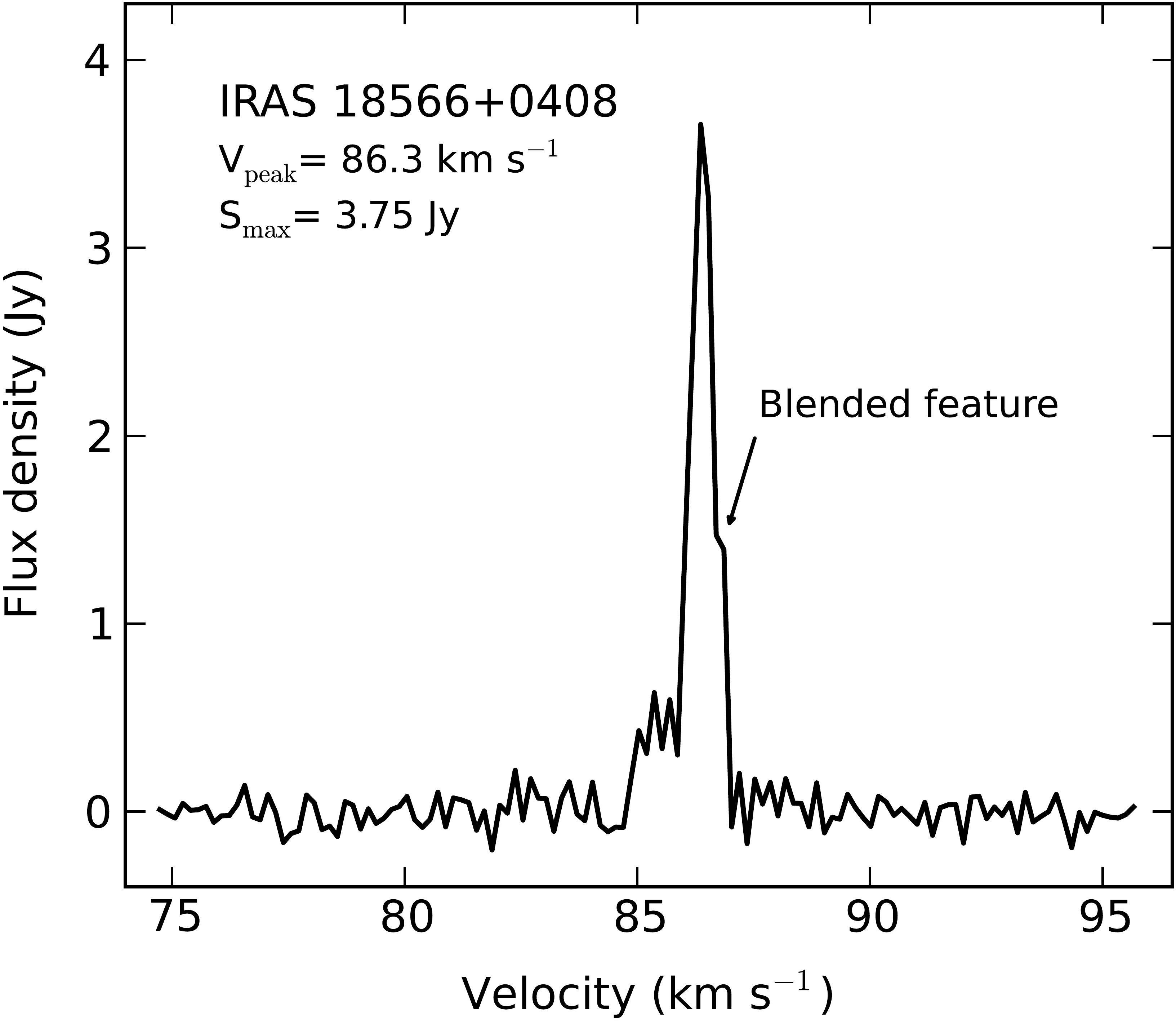}
\caption{Left: Spectrum of the strongest maser component found in our sample toward IRAS 18290$-$0924. Right: Example of a typical blended feature. We show the strongest maser component in source IRAS 18566+0408 which is blended with another feature at a nearby velocity (see Table \ref{table:masers}). \label{fig:spectra}}
\end{figure}

We detected 44 GHz methanol masers in 24 out of 56 fields (a 43$\%$ detection rate).  A total of 83 maser components were detected; for these masers we report the line parameters in Table \ref{table:masers}. Column 1 shows the IRAS name of the source. Column 2 lists a number associated to the maser component assigned by increasing Right Ascension. Columns 3 and 4 give the J2000 position of the peak emission obtained from a two-dimensional Gaussian fit to the peak maser channel. Column 5 is the LSR velocity of the peak emission channel. Column 6 is the flux density obtained from the 2D gaussian fit. Column 7 gives the full width of the line at zero intensity (FWZI) at a 4$\sigma$ level; in this case $\sigma$ is the channel-to-channel rms noise from the spectra. Column 8 shows the flux density integrated over velocity for channels above the 4$\sigma$ level of the spectra. Of the 83 maser components, there are 10 that overlap with a stronger maser component, causing a larger uncertainty in the reported parameters. We indicate these ten cases in Column 9. An example of an overlapping feature is shown in Figure \ref{fig:spectra}. The strongest maser in our sample showed a flux density of 62.73 Jy and was detected toward IRAS 18290$-$0924 (see Figure \ref{fig:spectra}).

To estimate the uncertainty in the flux density, we consider two contributing factors. First, the Q-band flux density calibration is limited to about 3\% accuracy due to pointing errors. The second is the statistical uncertainty of the flux bootstrapping for the phase calibrators reported by the task GETJY. The highest uncertainty reported is $\sim$ 6\%. Considering other sources of uncertainty in the VLA flux scale, we conservatively suggest an uncertainty of 15\%. We note that the flux densities reported in Table \ref{table:masers} were corrected for primary beam attenuation via the task PBCOR. We indicate correction factors greater than 2 in Column 9 of Table \ref{table:masers}. The uncertainty of the flux density is larger for these masers, and increases with the primary beam correction factor applied (i.e., the further the maser is outside the beam half-power point, the larger is the uncertainty). We used 3C147 and 3C286 as flux calibrators for the observing runs made during August and 3C48 and 3C286 for the observations taken in September. To set the flux density scales, we used the coefficients defined by Perley-Butler 2010 (Perley \& Butler 2013).

To estimate the positional accuracy we consider three sources of uncertainty. The first uncertainty is given by the VLA calibrator code; these range from 0\rlap.{$''$}002 to 0\rlap.{$''$}15 for calibrator codes A, B and C (see Table \ref{table:calibrator}). The second uncertainty depends on the image quality and was calculated using the expression reported by Condon et al. (1998): $\theta = \frac{1}{2} \theta_{\mathrm{syn}} \frac{\sigma_{\mathrm{rms}}}{\mathrm{I_{peak}}}$. We use the largest synthesized beam size and the weakest maser from each observing run to obtain an upper limit to the uncertainty; this contribution ranges from 0\rlap.{$''$}004 to  0\rlap.{$''$}095. An additional source of positional error is the phase transfer between calibrator and source. We estimated this positional uncertainty from the phase noise of the calibrators which ranged from 9$^\circ$ to 65$^\circ$. We use the expression $\mathrm{\theta = \frac{\sigma_\phi\theta_{\mathrm{syn}}}{2\pi\sqrt{n_p}}}$ where $\sigma_\phi$ is the phase rms, $\theta_{\mathrm{syn}}$ is the synthesized beam and $\mathrm{n_p}$ is the number of visibilities for each phase calibrator. This contribution ranges from 0\rlap.{$''$}0002 to 0\rlap.{$''$}0009. As the positional uncertainty from the phase noise was derived from observations of phase calibrators, then, for the maser sources, the positional error should be taken as a lower limit. We add in quadrature these three sources of uncertainty to obtain a conservative estimate of the positional error. For the observations made during August we estimate an uncertainty of 0\rlap.{$''$}1 and for the observing runs taken in September, we estimate a positional uncertainty of 0\rlap.{$''$}2.

\startlongtable
\begin{deluxetable}{ccccccccc}
\tabletypesize{\footnotesize}
\tablecolumns{9} 
\tablewidth{0pt}
\tablecaption{Class I methanol masers detected \label{table:masers}}
\tablehead{
\colhead{Source} & \colhead{Maser} &  \multicolumn{2}{c}{J2000 Peak Coordinates} & \colhead{Peak Velocity\tablenotemark{1}} & \colhead{S$_{\rm max}$} & \colhead{$\Delta$v} & {$\int {\rm S_\nu dv}$} & \colhead{Notes\tablenotemark{2}}    \\
\cline{3-4} 
\colhead{IRAS name}                  &
\colhead{Number}                     &
\colhead{$\alpha$($^{\rm h~ m~ s}$)}    & 
\colhead{$\delta$($^\circ$  $'~ ''$)} & 
\colhead{(km s$^{-1}$)}              &
\colhead{(Jy)}                       & 
\colhead{(km s$^{-1}$)}               & 
\colhead{(Jy km s$^{-1}$)}            & 
\colhead{}
 } 
\colnumbers
\startdata
05358$+$3543  & 1 & 05 39 11.37  & $+$35 45 08.4 &$-$18.4 & 0.84  & 0.49 & 0.20  &     \\  
              & 2 & 05 39 13.20  & $+$35 45 36.0 &$-$18.0 & 5.01  & 1.16 & 2.65  & 3.4 \\  
              & 3 & 05 39 13.48  & $+$35 45 43.0 &$-$16.6 & 24.72 & 0.99 & 14.91 & 5.8 \\  
18089$-$1732  & 1 & 18 11 51.29  & $-$17 31 23.8 & $+$31.5 & 17.70 & 3.31 & 10.82 &  \tablenotemark{a} \\ 
              & 2 & 18 11 51.39  & $-$17 31 24.6 & $+$32.5 & 1.87  & $-$    & $-$     &  \tablenotemark{a} \\
              & 3 & 18 11 51.41  & $-$17 31 22.2 & $+$33.8 & 1.52  & 1.66 & 0.37  & \\       
              & 4 & 18 11 51.51  & $-$17 31 21.6 & $+$33.6 & 1.55  & 2.32 & 1.51  & \\
              & 5 & 18 11 51.64  & $-$17 31 22.0 & $+$33.1 & 1.00  & 0.33 & 0.06  & \\
              & 6 & 18 11 51.70  & $-$17 31 23.8 & $+$32.4 & 4.39  & 1.49 & 2.09  & \\
              & 7 & 18 11 52.18  & $-$17 31 21.7 & $+$31.8 & 1.20  & 0.33 & 0.22  & \\
18102$-$1800  & 1 & 18 13 10.29 & $-$18 00 03.8 & $+$22.5 & 7.64  & 0.49 & 2.52   & 3.7 \\ 
              & 2 & 18 13 11.01 & $-$18 00 02.4 & $+$21.4 & 5.87  & 0.33 & 1.55   & 2.3 \\ 
              & 3 & 18 13 11.10 & $-$18 00 01.4 & $+$21.4 & 3.45  & 0.49 & 0.80   & 2.1 \\ 
              & 4 & 18 13 11.17 & $-$18 00 04.2 & $+$22.9 & 3.63  & 1.32 & 2.90   & 2.3 \\ 
              & 5 & 18 13 11.24 & $-$18 00 03.1 & $+$22.0 & 2.39  & 1.32 & 0.65   & 2.2 \\ 
              & 6 & 18 13 11.82 & $-$18 00 01.8 & $+$19.6 & 7.36  & 1.66 & 6.40   &  \\ 
18151$-$1208  & 1 & 18 17 57.12 & $-$12 07 34.7 & $+$33.2 & 1.34  & 1.32 & 0.98   &  \\ 
              & 2 & 18 17 58.20 & $-$12 07 29.4 & $+$33.7 & 2.94  & 0.66 & 1.12   &   \\ 
              & 3 & 18 17 58.56 & $-$12 07 17.0 & $+$32.1 & 13.83 & 0.99 & 7.08   &  \\ 
              & 4 & 18 17 58.58 & $-$12 07 18.8 & $+$33.6 & 5.71  & 0.49 & 1.91   &  \\ 
18182$-$1433  & 1 & 18 21 08.78  & $-$14 31 46.7 & $+$60.7 & 2.31   & 1.32 & 1.19  & \\ 
              & 2 & 18 21 08.90  & $-$14 31 46.8 & $+$60.4 & 1.33   & 1.32 & 0.69  & \\ 
              & 3 & 18 21 09.14  & $-$14 31 49.7 & $+$61.2 & 33.63  & 2.32 & 24.85 & \\ 
              & 4 & 18 21 09.85  & $-$14 31 46.7 & $+$60.2 & 13.90  & 1.16 & 7.77  & \\ 
18247$-$1147  & 1 & 18 27 31.80 & $-$11 46 00.5 & $+120.7$ & 6.53 & 1.66 & 2.66 & \\ 
18264$-$1152  & 1  & 18 29 14.18 & $-$11 50 27.4 & $+$43.6  & 3.82  & 1.82 & 1.82  & \\ 
              & 2  & 18 29 14.28 & $-$11 50 18.9 & $+$44.4  & 7.40  & 2.32 & 4.13  & \\ 
              & 3  & 18 29 14.33 & $-$11 50 27.4 & $+$45.2  & 1.12  & 0.49 & 0.33  & \\ 
              & 4  & 18 29 14.38 & $-$11 50 24.8 & $+$43.6  & 14.80 & $-$ & $-$    &  \tablenotemark{a}\\ 
              & 5  & 18 29 14.39 & $-$11 50 24.7 & $+$43.1  & 19.73 & 1.66 & 12.47  &  \tablenotemark{a}\\ 
              & 6  & 18 29 14.49 & $-$11 50 17.4 & $+$43.9  & 12.19 & 1.49 & 6.41  & \\ 
              & 7  & 18 29 14.57 & $-$11 50 23.8 & $+$41.1  & 1.09  & 0.49 & 0.37  & \\ 
              & 8  & 18 29 14.66 & $-$11 50 26.4 & $+$42.9  & 2.04  & 0.99 & 0.71  & \\ 
              & 9  & 18 29 14.77 & $-$11 50 22.3 & $+$42.7  & 2.01  & 2.15 & 1.36  & \\ 
              & 10 & 18 29 15.42 & $-$11 50 11.1 & $+$42.7  & 1.20  & 0.82 & 0.55  & \\ 
18290$-$0924  & 1 &  18 31 43.18 & $-$09 22 31.7 & $+$85.8  & 1.23  & 0.16 & 0.16  & 2.4 \\ 
              & 2 &  18 31 43.40 & $-$09 22 22.5 & $+$83.9  & 62.73 & 2.32 & 29.41 & \\ 
              & 3 &  18 31 44.20 & $-$09 22 12.6 & $+$82.8  & 1.17  & 0.49 & 0.40  & \\ 
18306$-$0835  & 1 &  18 33 22.95 & $-$08 33 35.3 & $+$76.4 & 2.86  & 0.99 & 1.31  & \\ 
              & 2 &  18 33 23.96 & $-$08 33 30.2 & $+$79.9 & 5.45  & 1.32 & 4.97  & 2.4 \\ 
              & 3 &  18 33 24.07 & $-$08 33 34.5 & $+$77.6 & 18.23 & 2.98 & 20.10 & 2.5 \\ 
18308$-$0841  & 1 &  18 33 32.49 & $-$08 39 08.5 & $+$76.6 & 0.82  & 0.66 & 0.27  & \\ 
              & 2 &  18 33 33.08 & $-$08 39 17.5 & $+$76.4 & 0.75  & 0.33 & 0.08  & \\ 
              & 3 &  18 33 33.21 & $-$08 39 14.7 & $+$76.4 & 14.93 & 2.49 & 10.42 &  \tablenotemark{a} \\ 
              & 4 &  18 33 33.16 & $-$08 39 13.6 & $+$75.8 & 6.20  & $-$    & $-$     &  \tablenotemark{a} \\ 
              & 5 &  18 33 33.24 & $-$08 39 15.7 & $+$76.9 & 6.49  & $-$    & $-$     &  \tablenotemark{a} \\ 
18310$-$0825  & 1 & 18 33 47.31 & $-$08 23 43.1 & $+$84.2 & 1.10 & 0.49 & 0.24 & \\ 
18337$-$0743  & 1 & 18 36 27.98 & $-$07 40 22.7 & $+$60.7 & 6.53 & 1.99 & 4.01 & \tablenotemark{a}  \\ 
              & 2 & 18 36 28.00 & $-$07 40 24.0 & $+$58.5 & 2.45 & $-$    & $-$ & \tablenotemark{a}  \\ 
18345$-$0641  & 1 & 18 37 16.53 & $-$06 38 34.4 & $+$95.9 & 1.30 & 1.32 & 0.89 &                     \\ 
              & 2 & 18 37 16.56 & $-$06 38 25.6 & $+$94.9 & 0.57 & 0.16 & 0.07 &                     \\ 
              & 3 & 18 37 16.89 & $-$06 38 29.7 & $+$95.4 & 7.74 & 2.98 & 6.58 & \tablenotemark{a}   \\ 
              & 4 & 18 37 16.94 & $-$06 38 30.7 & $+$97.0 & 1.66 & $-$    & $-$    & \tablenotemark{a}   \\ 
              & 5 & 18 37 17.30 & $-$06 38 31.5 & $+$96.2 & 1.98 & 2.49 & 2.24 & \tablenotemark{a}   \\ 
              & 6 & 18 37 17.21 & $-$06 38 30.5 & $+$94.5 & 0.45 & $-$    & $-$ & \tablenotemark{a}   \\ 
              & 7 & 18 37 17.33 & $-$06 38 34.4 & $+$97.2 & 0.66 & 0.66 & 0.18 &                     \\ 
              & 8 & 18 37 17.92 & $-$06 38 47.0 & $+$95.7 & 6.39 & 0.33 & 1.44 &                     \\ 
18488$+$0000  & 1 & 18 51 25.42 & $+$00 04 06.1 & $+$83.1 & 11.65 & 1.66 & 7.97 & \\ 
18517$+$0437  & 1 & 18 54 14.46 & $+$04 41 44.3 & $+$44.0 & 1.09 & 1.16 & 0.73 & \\ 
              & 2 & 18 54 14.74 & $+$04 41 42.7 & $+$43.4 & 4.19 & 1.16 & 2.70 & \\ 
18521$+$0134  & 1 & 18 54 40.79 & $+$01 38 05.5 & $+$77.8 & 0.36 & 0.16 & 0.02 & \\ 
              & 2 & 18 54 40.90 & $+$01 38 04.8 & $+$78.1 & 0.66 & 0.82 & 0.36 & \\ 
18530$+$0215  & 1 & 18 55 34.76 & $+$02 19 00.5 & $+$80.1 & 1.14 & 2.15 & 1.32 & \\ 
              & 2 & 18 55 34.90 & $+$02 18 58.6 & $+$79.1 & 0.94 & 0.66 & 0.24 & \\ 
18553$+$0414  & 1 & 18 57 53.34 & $+$04 18 16.9 & $+$12.5 & 1.47 & 0.99 & 0.70 & \\ 
18566$+$0408  & 1 & 18 59 09.23 & $+$04 12 22.3 & $+$86.8 & 1.04 & 0.33 & 0.26 & \\ 
              & 2 & 18 59 09.30 & $+$04 12 25.7 & $+$86.3 & 3.75 & 1.49 & 2.90 &\tablenotemark{a} \\ 
              & 3 & 18 59 09.20 & $+$04 12 25.4 & $+$86.8 & 1.73 & $-$   & $-$ &\tablenotemark{a} \\ 
              & 4 & 18 59 09.97 & $+$04 12 14.8 & $+$83.8 & 1.28 & 1.82 & 0.92 & \\ 
              & 5 & 18 59 10.17 & $+$04 12 16.3 & $+$87.0 & 1.09 & 0.66 & 0.40 & \\ 
20126$+$4104  & 1 & 20 14 25.17  & $+$41 13 36.2 & $-$2.3  & 5.07  & 2.32 & 3.92  & \tablenotemark{a}  \\  
              & 2 & 20 14 25.22  & $+$41 13 35.1 & $-$2.3  & 3.88  & $-$  & $-$   & \tablenotemark{a}  \\  
              & 3 & 20 14 25.30  & $+$41 13 40.6 & $-$3.1  & 2.55  & 0.49 & 0.53  & \\ 
              & 4 & 20 14 25.41  & $+$41 13 37.9 & $-$3.3  & 7.87  & 2.15 & 5.59  & \\ 
              & 5 & 20 14 26.71  & $+$41 13 29.8 & $-$4.4  & 2.81  & 0.82 & 1.33  & \\ 
20293$+$3952  & 1 & 20 31 11.97  & $+$40 03 12.1 & $+$6.1  & 9.58   & 1.66 & 4.95  & \\ 
              & 2 & 20 31 12.80  & $+$40 03 20.8 & $+$6.9  & 10.17  & 2.15 & 6.58  & \\ 
              & 3 & 20 31 12.80  & $+$40 03 24.0 & $+$4.8  & 14.58  & 1.32 & 8.06  & \\ 
23033$+$5951  & 1 & 23 05 24.56 & $+$60 08 09.4 & $-$54.4 & 17.81 & 1.16 & 8.30 & \tablenotemark{a}\\ 
              & 2 & 23 05 24.72 & $+$60 08 09.3 & $-$52.9 & 2.14  & $-$  & $-$ & \tablenotemark{a}\\ 
23139$+$5939  & 1 & 23 16 10.83 & $+$59 55 20.7 & $-$44.5 & 0.91 & 0.66 & 0.28    \\ 
23151$+$5912  & 1 & 23 17 21.30 & $+$59 28 50.6 & $-$54.7 & 3.67 & 0.33 & 0.84 & \\ 
              & 2 & 23 17 21.86 & $+$59 28 45.6 & $-$52.7 & 4.02 & 0.99 & 2.18 & \\ 
\enddata
\tablenotetext{1}{We report the channel velocity of the peak emission.}
\tablenotetext{2}{All flux densities were corrected for primary beam attenuation. We show the primary beam correction factors larger than 2 to indicate sources for which this correction may result in larger flux density uncertainties.}
\tablenotetext{a}{This maser component is blended with another feature at a nearby velocity. We were not able to determine the linewidth and integrated flux density precisely. We report the values for the stronger component.}
\end{deluxetable}

\subsection{Line properties}

\subsubsection{Linewidths}

\begin{figure*}[h!]
\begin{center}
\includegraphics[bb=17 33 447 370, width=100mm]{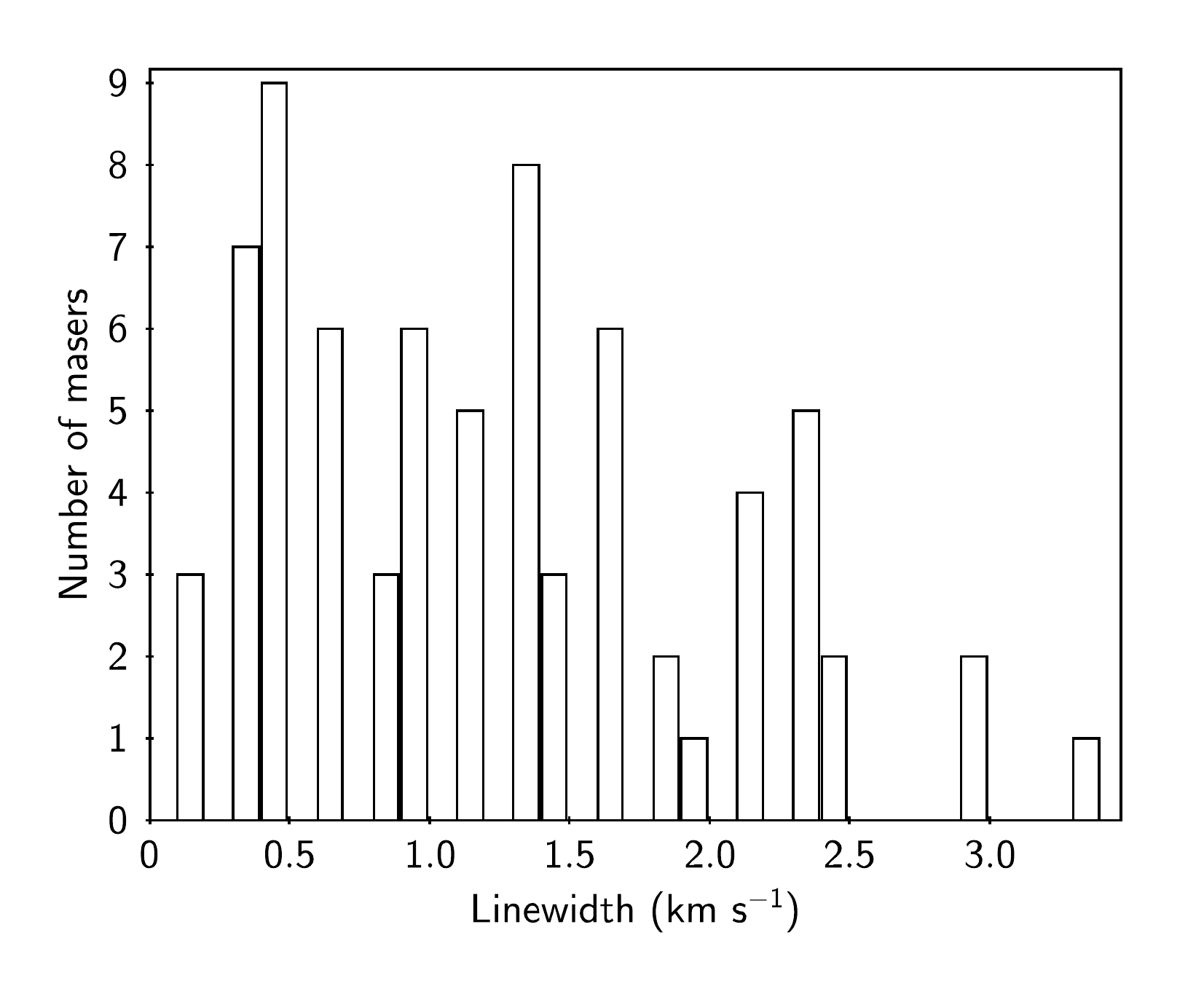}
\end{center}
\caption{Histogram of the maser linewidths in our sample. The first bin (with 3 masers) corresponds to components with linewidth equal to the spectral resolution.}
\label{lwh}
\end{figure*}

In Figure \ref{lwh} we show a histogram of the maser linewidths measured in our survey. The linewidths range from 0.16 km s$^{-1}$ (the channel width) to 3.31 km s$^{-1}$. We caution that components with linewidths greater than $\sim$ 1 km s$^{-1}$ may occur because of overlapping weak maser features (as reported in Table \ref{table:masers}) or because of a thermal (i.e., non-maser) contribution to the emission.

 The observations reported here were made in a similar manner to those of Paper III, using the VLA D-configuration with a spectral resolution of 0.16 km s$^{-1}$ and a velocity coverage of 21 km s$^{-1}$. Therefore, it is reasonable to make a direct comparison between the linewidth distributions obtained in the two surveys. We find very similar linewidth ranges with the difference that this work does not show a clear peak of the distribution whereas Paper III found a peak at 0.35 km s$^{-1}$.  The median linewidth of the present data is 1.16 km s$^{-1}$, similar to the 1.0 km s$^{-1}$ found in Paper III.

\subsubsection{Relative velocity distribution}

\begin{figure*}[h!]
\begin{center}
\includegraphics[bb=17 33 447 350,width=100mm]{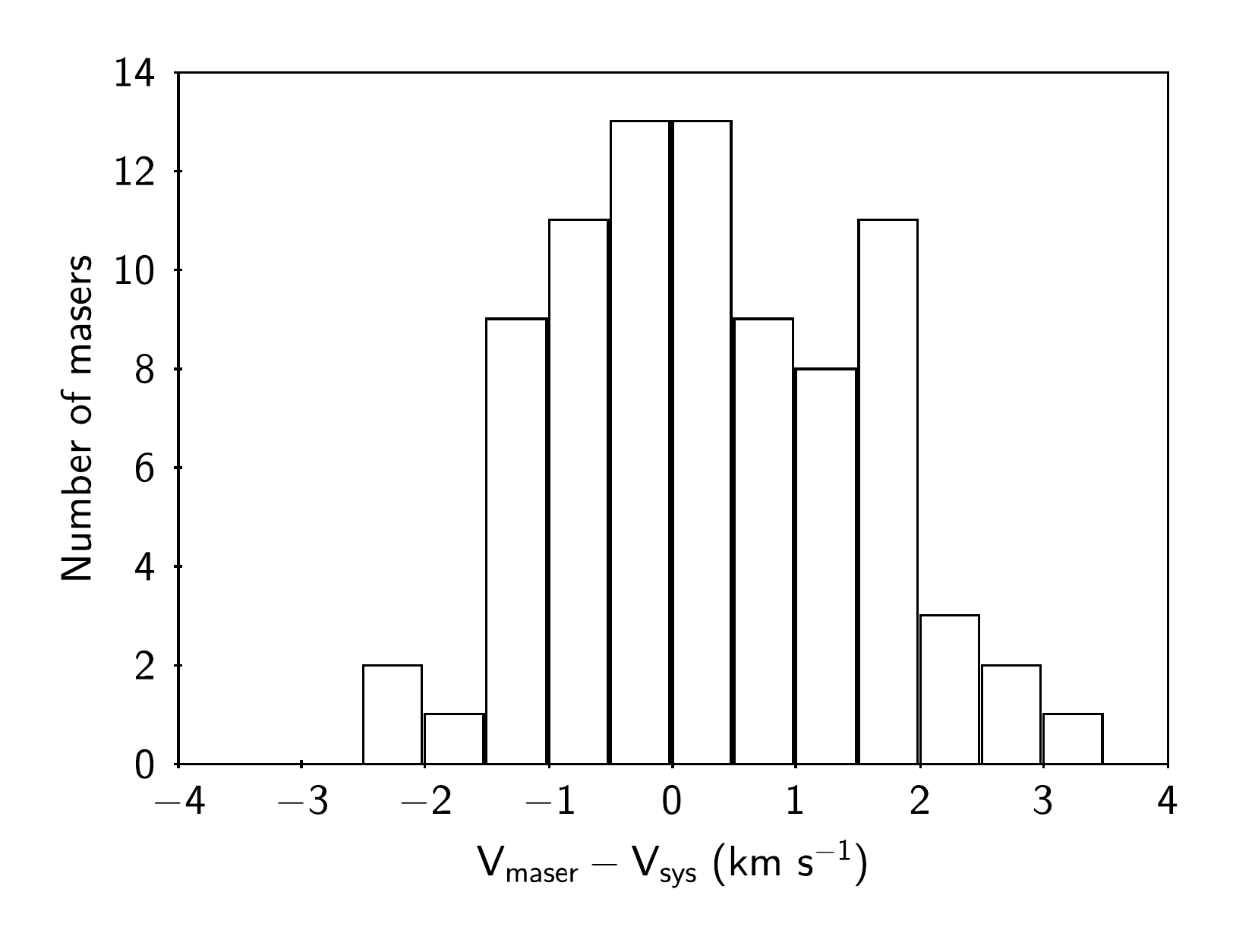}
\end{center}
\caption{Histogram of the maser$-$cloud relative velocity distribution.}
\label{vth}
\end{figure*}

In Figure \ref{vth} we show a histogram of the maser velocities relative to the ambient molecular cloud. We use the systemic cloud velocities reported by S02 from CS(2$-$1) observations and plot the difference between the maser velocity and the cloud velocity. The relative velocities range from $-$2.5 to +3.1 km s$^{-1}$; the mean value of the distribution is 0.22 km s$^{-1}$ with a standard deviation of 1.22 km s$^{-1}$. The total range of velocities is less than 5.6 km s$^{-1}$. The statistics from Paper III are similar to ours: their relative velocities range from $-$2.7 to +3.8 km s$^{-1}$ with a mean of 0.09 km s$^{-1}$ and a standard deviation of 1.27 km s$^{-1}$. The median value of both distributions is 0.2 km s$^{-1}$. Similar results were obtained by Jordan et al. (2015). Their distribution shows a total velocity range less than 9 km s$^{-1}$ with a mean value of 0.0$\pm$0.2 km s$^{-1}$ and a standard deviation of 1.5$\pm$0.1 km s$^{-1}$.

Although our histogram peaks near zero, there is a small but discernible redshifted asymmetry (see Figure \ref{vth}). This tendency was also noted in the velocity histogram of Paper III.

To determine if the redshifted trend is real, as suggested by inspection of the velocity distributions from this paper and Paper III (see our Figure \ref{vth} and their Figure 2), we performed the D'Agostino-Pearson test for normality on both samples to check for deviations from a normal distribution based on the skewness and kurtosis. The null hypothesis to test is that the relative velocity distribution comes from a normally distributed population. The normality test reports a p-value which is a measure of the significance level of the test. A p-value higher than 0.05 would be the criterion to accept the null hypothesis that the data have a normal distribution. We obtained a p-value of 0.44 and 0.61 for the velocity distributions of this paper and Paper III, respectively. Thus, the normality test suggests a 44\% and 61\% probability that both datasets are consistent with a normal distribution. Although we cannot confirm the redshifted tendency in our data, we consider this point to be an open question.

Kirsanova et al. (2017) compared 6.6 GHz maser velocities with systemic cloud velocities traced by CS(2$-$1) toward 24 high-mass star-forming regions in the Perseus spiral arm. They find a predominance of redshifted masers and suggest this is related to large-scale galactic motions.

\subsection{Comments on individual sources}

In Table \ref{table:refs}, we summarize single-dish and interferometric observations of some common star formation sign posts made toward the 24 sources with 44 GHz maser detections. We consider seven sources for special comments because they have additional information available in the literature to understand their nature in relation to the 44 GHz masers. Three-color images from Spitzer/IRAC of all regions with 44 GHz masers are shown in Figure \ref{irac}\footnote{This research has made use of SAOImage DS9, developed by the Smithsonian Astrophysical Observatory.}.

\subsubsection{IRAS 05358$+$3543 \label{sec:05358}}

This source is part of the star-forming complex S233IR located at a distance of 1.8 kpc (Snell et al. 1990). S233IR contains two young embedded clusters in different evolutionary states separated by 0.5 pc. They are labeled as the NE and SW clusters with estimated ages $\lesssim$ 2 Myr and 3 Myr, respectively (Porras et al. 2000). Several star formation tracers are present in the NE cluster: H$_2$O and OH masers (Tofani et al. 1995; Argon et al. 2000; Beuther et al. 2002c), intense Class II 6.6 GHz methanol masers (256 Jy; Menten et al. 1991; Szymczak et al. 2000) and at least three highly-collimated jets revealed from shocked H$_2$ emission (Qiu et al. 2008; Porras et al. 2000). Three millimeter sources and possibly four molecular outflows are found in this cluster (mm1, mm2 and mm3; outflows A, B, C and D: Beuther et al. 2002d). Outflow A is a large-scale ($\sim$ 1 pc) highly-collimated CO outflow possibly powered by mm1; outflow B is a high-velocity CO outflow and outflow C is observed mainly in SiO emission. In marked contrast, the SW cluster --- which harbors the IRAS source --- does not show any of the star formation activity described above. No radio continuum emission at 3.6 cm was detected toward either cluster (S02).

Our observations were centered on the IRAS source in the SW cluster where one maser spot was detected; the NE cluster is outside our primary beam although two maser spots were detected here (see Figure \ref{irac}). The largest primary beam correction factors reported in Table \ref{table:masers} were applied to this source. In total, three methanol maser spots were detected; the one detected toward the SW cluster is coincident with SiO emission from the southern lobe of outflow C and also with an extended green feature that appears emanating from the SW cluster in a NW-SE direction. The other two in the NE cluster are either coincident with the inner knots of outflow A or with the tip of the south-eastern lobe of outflow B. One of these two masers is also coincident with H$_2$ emission (knot N4A: Porras et al. 2000; bow shock 1: Varricatt et al. 2010; Beuther et al. 2002d). In contrast to the Class II, H$_2$O and OH masers which are projected near the center of the NE cluster (Porras et al. 2000), the Class I masers are located at the borders of the star formation regions.

 Litovchenko et al. (2011) made single-dish observations of 44 GHz masers in this source. Their pointing was centered toward the NE cluster and did not detect emission. Their primary beam covers the complete area where we detect two bright masers of 24.72 Jy and 5.01 Jy. Their nondetection suggests variability.

\subsubsection{IRAS 18089$-$1732}

This source is common to the S02 sample and the ``High'' sub-sample of Molinari et al. (1996). Beuther et al. (2004a) favor the near kinematic distance of 3.6 kpc corresponding to a luminosity of $\sim$ 32,000 L$_\odot$ (S02). A compact core oriented in the north-south direction was found through interferometric continuum observations at 1.3 mm and 3.6 cm (Beuther et al. 2004a; Zapata et al. 2006). Emission from SiO(5$-$4) reveals a molecular outflow with both the red and blue lobes oriented to the north of the core (Beuther 2004a). Sensitive VLA observations at 7 mm resolve the core into two components: sources ``a'' and ``b'' (Zapata et al. 2006). Zapata et al. interpret source ``a'' as a thermal jet that is driving the SiO(5$-$4) outflow and, source ``b'' as an optically thick HII region. H$_2$O, OH and Class II 6.6 GHz methanol masers are projected inside the 3.6 cm continuum source (Argon et al. 2000; Beuther et al. 2002c).

Six of the seven 44 GHz masers are distributed in an arc-like feature about 8$''$ (0.1 pc) to the north of the compact core and the star formation indicators mentioned above (see Figure \ref{irac}). We speculate that this arc feature could be tracing the shocked gas from the SiO(5$-$4) outflow lobes. The lobes extend northward about 5$''$ from the millimeter peak with a 20$^\circ$ position angle; the 44 GHz masers are located at the northern edge of the outflow. The 44 GHz maser velocities are within a few km s$^{-1}$ of the systemic velocity (+33.8 km s$^{-1}$ from S02), while the SiO outflow has a velocity range from +26 to +41 km s$^{-1}$. These masers appear at the interface between the outflowing gas and the ambient material. Fontani et al. (2010) report single-dish detections of both 44 and 95 GHz Class I maser emission that closely matches the velocities of the 44 GHz masers that we detect. The isolated CH$_3$OH maser to the east of the arc-like feature is close to (but not coincident with) a weak green excess in the Spitzer image.

\subsubsection{IRAS 18102$-$1800}

The near and far distances reported by S02 for this source are 2.6 kpc and 14.0 kpc corresponding to 10$^{3.8}$ and 10$^{5.3}$ L$_\odot$, respectively. The Spitzer image shows an extended source with an 8 $\mu$m excess --- which hosts the IRAS source --- and a compact object $\sim 20''$ to the SW of the IRAS position. IRAS 18102$-$1800 has a bright 3.6 cm continuum source (44 mJy) detected with the VLA.  Class II methanol masers are located at the peak position of the SW compact source (Beuther et al. 2002c). An extended millimeter core is located $5''-7''$ south of the Class II masers (see Fig. 1 in Beuther et al. 2002c). No water masers were detected (S02).

Our observations were centered on the IRAS position but the majority of the 44 GHz methanol masers are distributed along the SW compact source which is at the edge of the VLA primary beam (see Figure \ref{irac}). Large primary beam correction factors were applied to this source (see Table \ref{table:masers}). The 44 GHz masers lie roughly along an east-west line of length $\sim$ 0.3 pc (1.6 pc) for a distance of 2.6 kpc (14.0 kpc). The absence of centimeter continuum emission --- but the presence of Class II methanol masers --- suggests that the compact SW object is in a younger evolutionary state than the IRAS source. If so, the indication is that 44 GHz appear preferentially toward the younger source. Alternatively, the masers at 44 GHz may trace an outflow powered by a protostar near the millimeter core.

\subsubsection{IRAS 18182$-$1433}

A distance of 3.6 kpc has been adopted for this source (Moscadelli et al. 2013). The IRAS source is formed by two mid-IR objects separated by 10$''$ along the NW-SE direction (De Buizer et al. 2005). VLA observations at 7 mm, 1.3 cm and 3.6 cm detected one millimeter source (source ``a'') and two centimeter sorces (sources ``b'' and ``c''; Zapata et al. 2006). Sources ``a'' and ``b'' are coincident with the NW mid-IR object while source ``c'' coincides with the stronger SE object. Class II CH$_3$OH, H$_2$O and OH masers are located toward the region of sources ``a'' and ``b'' (Walsh et al. 1998; Forster \& Caswell 1999; Beuther et al. 2002c; Sanna et al. 2010). Multiple CO and SiO outflows seem to emanate from the location of sources ``a'' and ``b'' (Beuther et al. 2002b; 2006). None of these star formation tracers are seen toward source ``c''. All these data suggest that the NW source is younger and more embedded than the SE source (De Buizer et al. 2005).

Our observations were pointed toward the position reported by S02 which is offset from the IRAS position by 19\rlap.{$''$}5 (see Figure \ref{irac}). We detect 4 Class I methanol masers which are distributed in the region around the NW sources ``a'' and ``b'' but no masers were found toward the SE source ``c''. The 44 GHz masers are located at the edges of the green excess seen in the 4.5 $\mu$m band. The suggestion in this case, is that 44 GHz masers tend to favor the younger objects in the region.

\subsubsection{IRAS 18264$-$1152 \label{sec:18264}}

S02 report near and far kinematic distances of 3.5 and 12.5 kpc, respectively, with luminosities of 10$^4$ L$_\odot$ and 10$^{5.1}$ L$_\odot$. Class II CH$_3$OH, H$_2$O and Class I 95 GHz masers are reported in the region (S02; Beuther et al. 2002c; Chen et al. 2011). Multiple molecular outflows were revealed from interferometric and single-dish observations: a high-velocity SiO outflow ($\Delta$v $\sim$ 60 km s$^{-1}$) and a CO outflow oriented in an east-west direction (Beuther et al. 2002b; Qiu et al. 2007) also traced by H$_2$ emission (Varricatt et al. 2010). Although undetected at 3.6 cm by S02, more sensitive observations detected emission at the 1 mJy level (Zapata et al. 2006; Rosero et al. 2016). This source is resolved into three components at 1.3 cm and 7 mm: called sources ``a'', ``b'' and ``c''. Zapata et al. suggest an optically thick HII region or dust emission from a core and disk for source ``a'', a thermal jet or a partially optically thick HII region for source ``b''; and an optically thick HII region or dust emission from a core and disk for source ``c''.

This source shows the highest level of 44 GHz maser activity in our survey. We report 10 maser components; 9 of them are located in the general vicinity of the YSO and other star formation indicators. The remaining maser spot is isolated, some 10$''$ to the NE. Two of the 9 masers are projected against the 3.6 cm emission and near the peak position of H$_2$O and Class II methanol masers. IRAS 18264$-$1152 has been classified as an EGO by Cyganowski et al. (2008), indicating the presence of shocked gas.

 None of these maser components with flux densities ranging from 1.09 to 19.73 Jy (see Table \ref{table:masers}) were detected by the single-dish observations of Litovchenko et al. (2011) which suggests maser variability.

\subsubsection{\label{18517} IRAS 18517+0437}

This source is located at a distance of 2.9 kpc and has a luminosity of $\sim$ 13000 L$_\odot$ (S02). Two mid-IR objects located with a NE-SW orientation contribute to the IRAS source. A third object with a green bipolar structure consistent with an EGO is located some 10$''$ east of the NE source. Intense Class II 6.6 GHz methanol masers were found by single-dish observations (279 Jy; Schutte 1993; Szymczak 2000) as well as H$_2$O 22 GHz masers (Brand et al. 1994; Codella et al. 1996; S02) but their precise positions are unknown, making it difficult to compare their positions with those of the other objects in the field. Thermal free-free emission at 3.6 cm was not detected by S02 but recent VLA observations detected three weak sources at 6 cm (Rosero et al. 2016; see Figure \ref{irac}). Source ``A'' as labeled by Rosero et al. appears to power the EGO. Although CO line wings were detected by S02, no molecular outflows were found by the CO observations of Zhang et al. (2005). This source was also observed in the survey of Paper III; the line parameters reported are consistent with our findings within the uncertainties.

We find two 44 GHz methanol maser spots located at the borders of the green emission, consistent with the idea that EGOs trace shocked gas where methanol masers are excited. 

\subsubsection{IRAS 20293$+$3952 \label{sec:20293}}

\begin{figure*}[h!]
\begin{center}
\includegraphics[width=90mm]{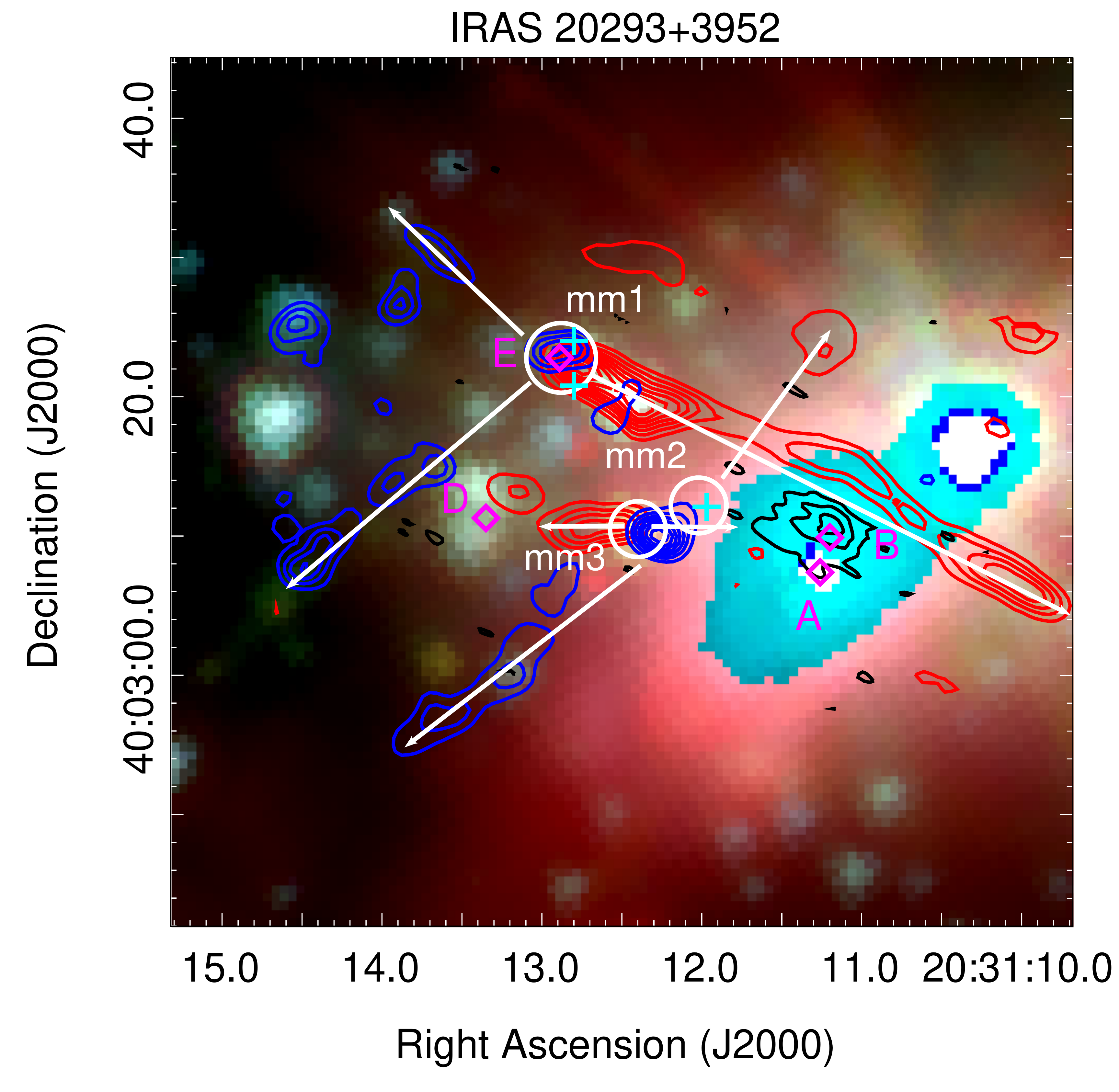}
\end{center}
\caption{ Three-color image from Spitzer (blue = 3.6 $\mu$m, green = 4.5 $\mu$m; red = 8 $\mu$m). The red and blue contorns represent the CO molecular outflows detected by Beuther et al. (2004b). The white arrows indicate the direction of the molecular outflows. The black contours trace the UCHII region from Beuther et al. (2004b).  The white circles symbolize the three millimeter sources (mm1, mm2, mm3) reported by Beuther et al. (2004b). The magenta diamonds indicate the peak position of the four compact centimeter sources (A, B, D, E) detected by Rosero et al. (2016). The cyan crosses represent the three 44 GHz maser spots reported in this work.}
\label{fig:20293}
\end{figure*}

This region is a typical source from our sample which hosts multiple star formation sites (see Figure \ref{fig:20293}). It is located at a distance of 2 kpc (Beuther et al. 2004b). The IRAS source is associated to an UCHII region which dominates the luminosity of the region $\sim$ 6300 L$_\odot$ and has an estimated ionizing photon rate corresponding to a B1 zero-age main-sequence star (Beuther 2004b; Palau et al. 2007). IR studies have reveal two regions, IRS 1 and IRS 2. IRS 1 is a possible binary system (Palau et al. 2007), in which the northern and fainter source is associated with the UCHII region and the whole system is surrounded by a ring of H$_2$ emission (Kumar et al. 2002; Varricatt et al. 2010). Approximately $20''$ north-east of the UCHII region there is a dense cloud of $\sim$ 250 M$_\odot$ and $\sim$ 0.5 pc of size that hosts a number of YSOs in different evolutionary stages from starless cores to (proto)stars (mm1, mm2, mm3, BIMA 3, BIMA 4, etc; Beuther et al. 2004b; Palau et al. 2007). The millimeter source mm1 is an intermediate-mass protostar of 4 M$_\odot$ while mm2 and mm3 are low-mass protostars with estimated masses around 1 M$_\odot$. All of these millimeter sources are associated with four highly-collimated outflows identified from CO(2$-1$) and SiO($2-1$) interferometric observations (Beuther et al. 2002b; 2004b; outflows A, B, C and D). Outflows A and B emanate from mm1; outflow C from mm3, and outflow D from mm2 (see Figure \ref{fig:20293}). Weak radio continuum sources were detected by very sensitive VLA observations at 6 cm and 1.3 cm, see Figure \ref{fig:20293} (sources A, B, C, D and E; Rosero et al. 2016). Particularly, source B is a weak and compact object projected inside the UCHII region which is similar to that found in other UCHII regions such as W3(OH) and M8 (Dzib et al. 2013; Masqu\'e et al. 2014). The peak position of source E is coincident with the position reported for the millimeter source mm1 and also with a water maser reported by Beuther et al. (2002c).

Figure \ref{fig:20293} shows contours of CO emission tracing the molecular outflows and the UCHII region overlaid with the three 44 GHz masers detected. Two of the three 44 GHz masers overlap with mm1 and radio source E while the other maser is very close to the peak position of mm2. The methanol masers toward mm1 arise near the base of outflows A and B rather than at the interaction region of the outflow lobes with the ambient medium. A green feature seems to extend in the same direction as outflow B but no masers were found along the outflowing gas. The excitation of the maser close to mm2 is not clear; it is possible that the excitation may be related to the outflows from the low-mass protostars mm1 and mm2. Only a few cases of Class I masers in low-mass sources have been reported (Kalenskii et al. 2010); this maser may be a similar case. Because this maser is close to an H$_2$ knot from the expanding ring-like structure, an alternative explanation is that this maser traces shocked gas from the interaction between the expanding UCHII region and the ambient molecular environment.

\section{Discussion \label{sec:discuss}}

\subsection{Comparison with previous surveys}

 Several sources appear multiple times in Papers I-IV. Paper I observed 44 massive star-forming regions in different evolutionary states, taken from the catalogs of Bachiller et al. (1990), Haschick et al. (1990), Slysh et al. (1994) and Molinari et al. (1996). Three of the sources in Paper I couldn't be imaged and so were re-observed in Paper II to complete the previous survey and provide accurate positions and line parameters. Paper III presented a study of 69 YSOs taken from the catalog of high-mass protostellar candidates of Molinari et al. (1996). Thirteen of these 69 YSOs were previously studied in Paper I. The criteria used to select the sample of Paper I were different to those used for the subsequent papers. Paper I consists of a heterogenous collection of sources while the sample of Paper III satisfies specific selection criteria and so was selected systematically. The overlap of 13 YSOs between Papers I and III is merely a consequence of the different goals of each project.

The present paper (IV) consists of 56 HMPOs taken from the S02 catalog. One source, IRAS 20126+4104, is common to the samples of Papers I and III. Another source, IRAS 18517+0437, is common to Paper III. These two sources were observed in a similar manner in each survey, using the VLA D-configuration with a spectral resolution of 0.16 km s$^{-1}$ and a velocity coverage of 21 km s$^{-1}$. Thus, it is reasonable to make comparisons between the maser line properties obtained in the different surveys to check the consistency of the observations and as a test of variability.

For 20126+4104, we detect five maser components. Components 1$-$4 (see Table \ref{table:masers}) are clustered $\sim 10''$ to the northwest of the IRAS source while component 5 is located $\sim 10''$ to the southeast of the IRAS source (see Figure \ref{irac}). For all 5 maser components, the coordinates and velocities coincide within the uncertainties for all three observations. The maser fluxes are similar between Papers I and III; in this work, however, we find somewhat higher fluxes. The peak flux densities of components 1 to 4 are almost a factor of 2 larger than those reported in Papers I and III. For example, component 4 has the largest peak flux density of 7.87 Jy while Papers I and III measured peak fluxes of 3.91 Jy and 4.24 Jy, respectively. In contrast, the peak flux density and linewidth of the isolated component 5 is consistent with the two previous observations, i.e., it remained nearly the same. The data were treated in a similar manner in each of the three surveys with the exception that we apply primary beam correction. Nevertheless, the maser locations are near the pointing center, consequently the correction factors are small (less than 8\%) which is insufficient to explain the difference in measured fluxes. A common amplification factor to the maser peak fluxes of the NW group and the absence of this amplification in the isolated maser in the SE suggests that some physical event ocurred between March 2007 and August 2008, affecting only the NW masers.

Class I masers are known to be collisionally pumped, so it is plausible to suppose a shock front emerging from the HMPO as the cause of the NW maser group brightening. We consider maser components 3 and 4 to define the width of a spherical shell surrounding the IRAS source which is located at a distance of 1.64 kpc (Moscadelli et al. 2011). The radial separation between IRAS 20126+4104 and maser components 3 and 4 is 12.98$''$ and 10.10$''$, respectively. This corresponds to a shell width of about 2.9$''$ (0.023 pc). We use the 15 month time separation between the observations as an upper limit to the travel time for such a shock moving along the shell. This implies a lower limit to the shock velocity of about 18,300 km s$^{-1}$ which is unrealistically high. An alternative possibility is that the putative shock originated from a different location.

For 18517+0437, we detect two maser components. The positions, flux densities and linewidths are consistent with the parameters reported in Paper III within the uncertainties. The brightest maser we measured has a peak flux density of 4.19 Jy (component 2) while Paper III reported a maser peak flux of 3.32 Jy. The other maser we detect has a flux density of 1.09 Jy (component 1) while Paper III reported a flux of 1.28 Jy. A weak third component was reported in Paper III but we did not detect it. The flux density variation between both Papers is because we apply primary beam correction and Paper III did not. Components 1 and 2 are near the edges of the VLA primary beam (see Figure \ref{irac}), so the correction factors here become significant. In fact, for component 2, we measured exactly the same flux density than Paper III (3.32 Jy) but the 4.19 Jy reported is the result of applying a correction factor of 26\% of the measured flux. This source is discussed in detail in Section \ref{18517}.

A single-dish survey of 59 sources was made by Litovchenko et al. (2011) to search for 44 GHz methanol masers; 22 of the 59 sources were among our sample. They used a broad-bandwidth spectrometer with a spectral resolution of 0.18 km s$^{-1}$, similar to our resolution. They did not report the detection limit of their observations but their spectra indicate a 1$\sigma$ noise level of about 1 Jy. They detect maser emission in 6 of the 22 sources; the line parameters reported are similar to our results. However, they report 4 sources as nondetections while we detect maser activity (05358+3543, 18264$-$1152, 18566+0408, 23139+5939). Their single-dish pointing centers are the same as our VLA pointings, except for source 05358+3543. Despite this pointing difference, for IRAS 05358+3543 we detect 2 masers of 24.72 Jy and 5.01 Jy in a region well-covered by their single-dish primary beam. Their nondetection suggests variability. For IRAS 18264$-$1152, we detect 10 maser components with flux densities ranging from 1.09 to 19.73 Jy but none of them were detected by Litovchenko et al., again suggesting variability. These two sources are discussed in detail in Sections \ref{sec:05358} and \ref{sec:18264}. For IRAS 18566+0408, we detect five maser features with fluxes from 1.04 to 3.75 Jy but were not detected by Litovchenko et al., however, these may have been below their detection limit. For IRAS 23139+5939, we detect one maser with a flux density of 0.91 Jy which was not detected by their single-dish observations. As in the previous case, the maser we detect is below their sensitivity limit.

\subsection{Relation with molecular outflows \label{sec:egos}}

The GLIMPSE survey, performed with the Spitzer Space telescope\footnote{This work is based [in part] on observations made with the Spitzer Space Telescope, which is operated by the Jet Propulsion Laboratory, California Institute of Technology under a contract with NASA.}, is a powerful tool to identify high-mass YSO with outflow activity (Benjamin et al. 2003; Churchwell et al. 2009). Extended and enhanced emission at 4.5 $\mu$m is commonly known as ``extended green objects'' (EGOs; Cyganowski et al. 2008; 2009) or ``green fuzzies'' (Chambers et al. 2009) as they appear green in the three-color composite images (3.6 $\mu$m in blue, 4.5 $\mu$m in green and 8 $\mu$m in red; Fazio et al. 2004). The 4.5 $\mu$m band contains H$_2$ lines and the CO fundamental band (De Buizer \& Vacca 2010) and hence strong emission in this band is an indicator of shocked gas. Although the exact nature of EGOs is still uncertain, it is likely that they arise in shocks where protostellar outflows collide with the ambient gas.

Nearly all of the sources in our sample were observed by the GLIMPSE survey and the majority of them are associated with CO line emission related to molecular outflows (Beuther et al. 2002b). Class I methanol masers are collisionally pumped and it has been suggested that they are excited at the interface where outflows encounter the molecular ambient medium (Plambeck \& Menten 1990). To investigate this assertion, we used the GLIMPSE survey to search for correlations between 44 GHz masers and EGOs.

We present three-color images in Figure \ref{irac} along with the peak positions of our 44 GHz methanol maser detections for the 24 fields. Three of the 24 fields were catalogued as EGOs by Cyganowski et al. (2008; 18182$-$1433, 18264$-$1152, and 18566+0408). The images reveal nine more fields with extended green emission with morphology similar to EGOs (05358$+$3543, 18151$-$1208, 18247$-$1147, 18290$-$0924, 18345$-$0641, 18488+0000, 18517+0437, 20126$+$4104 and 20293$+$3952). Except for 18517+0437, the other three sources are well-correlated with CO emission which traces the molecular outflows (Beuther et al. 2002b; 2002d; Zhang et al. 2005). The masers appear in some cases at the base of the outflows while in others they appear at the outflow lobes.

Two more sources with 44 GHz masers (18102$-$1800 and 18488$+$0000) and bright centimeter emission have a special behavior. In these two cases each field is dominated by an extended red feature which is also traced by bright centimeter emission with a morphology typical of a cometary UCHII region. Next to the dominant feature, both fields show a compact green object. The masers are preferentially located toward the compact green feature, and not toward the centimeter sources.

For sources 23033$+$5951 and 23139$+$5939 Spitzer images are not available; instead, we use images from the Wide-field Infrared Survey Explorer (WISE; Wright et al. 2010)\footnote{This publication makes use of data products from the Wide-field Infrared Survey Explorer, which is a joint project of the University of California, Los Angeles, and the Jet Propulsion Laboratory/California Institute of Technology, funded by the National Aeronautics and Space Administration.}. IRAS 23033+5951 also presents extended green emission in the 4.6 $\mu$m band but, due to the lower angular resolution of WISE compared to IRAC, no relationship with the masers can be established.

\subsection{On the classification between HMPOs and UCHII regions \label{sec:classif}}

 We discuss here the distinction made in Paper III between HMPOs and UCHII regions. In that paper, the objects were classified based on whether centimeter continuum emission from ionized gas (at a detection limit of $\sim$ 1 mJy) was found near the IRAS source (less than 40$''$). Regions with no centimeter emission were referred to as HMPOs and those with centimeter emission as UCHII regions. Given this classification, the conclusion in Paper III was that methanol masers are more common toward the more-evolved sources of their sample, i.e., toward the UCHII regions.

In this Paper, we avoid the HMPO/UCHII classification for three reasons: 1) it is ambiguous, since the free-free thermal emission may have other explanations besides a UCHII region. Different phenomena can give rise to such emission, for example, shocks produced by the interaction of thermal jets with surrounding material (Rodr\'iguez et al. 2012), stellar winds (Carrasco-Gonzalez et al. 2015), etc. 2) As more sensitive continuum observations become available, weaker continuum emission will be detected and therefore, almost all sources will be classified as UCHII regions, regardless of the nature of the emission, and 3) because massive stars form in cluster environments, HMPO and UCHII regions may co-exist within a single star formation region. To classify the entire region as ``more-evolved'' owing to the presence of UCHII region could be misleading --- especially if the masers are clearly associated with a HMPO and not the UCHII region.

S02 defined their HMPOs candidates as isolated young objects in evolutionary stages prior to the formation of a UCHII regions and/or without more-evolved objects (with photo-ionized gas) in their near vicinity. They based this definition on the fact that the objects were undetected in 5 GHz single-dish surveys of free-free thermal emission at a level of $\sim$ 25 mJy (Gregory \& Condon 1991; Griffith et al. 1994; Wright et al. 1994). Once they selected a sample of 69 HMPOs, they made follow-up VLA 3.6 cm observations at a level of 0.1 mJy and found that 40 of the previously undetected 69 sources show continuum emission above their detection limit of 1 mJy. Recently, with the improved VLA continuum sensitivity, deeper 1.3 cm and 6 cm observations with $\sim 10 \mu$Jy sensitivity have revealed multiple and even weaker and more compact radio sources in 23 of the 29 non-detections from S02 (Rosero et al. 2016). 

In this paper, we observed 56 of 69 sources from S02.  At the time when S02 selected their sources and following the Paper III classification, we would have had 0 UCHII regions and 56 HMPOs. Our detection rate would have been 43\% toward the younger sources and 0\% toward more-evolved sources. After their follow-up VLA observations, S02 detected centimeter continuum emission ($\geq$ 1 mJy) in 34 of the 56 fields while 22 were undetected ($<$ 1 mJy). The more sensitive images at the $\mu$Jy level ($\sim 3-50$ $\mu$Jy) toward 15 of the 22 non-detections revealed weak and compact centimeter emission (Zapata et al. 2006; Rosero et al. 2016). Following the Paper III classification we would have a total of 49 UCHII regions and 7 HMPOs. We detect maser emission in 22 of the 49 UCHII regions and in 2 of the 7 HMPOs. In this case, our conclusion would be a 45\% (22/49) detection rate of methanol masers toward the more evolved objects (UCHII regions) and a 29\% (2/7) detection rate toward the younger sources (HMPOs) of our sample, i.e.; masers would appear more common toward the UCHII regions. As pointed out before, as more sensitive observations become available, more centimeter sources will appear, further increasing the fraction of sources classified as UCHII regions regardless of the nature of the ionized gas. Moreover, a spectral index analysis of the 15 weak sources indicates that in almost all cases the emission comes from thermal jets, suggesting that these compact and weak sources are in a pre-UCHII region phase, so, they should not be counted as UCHII regions (Rosero et al. 2016).

A final factor to consider is that many of the 56 fields observed have multiple objects in different evolutionary stages, as discussed in Section \ref{sec:20293}. This was revealed from IR, millimeter and centimeter observations (S02; Beuther et al. 2002a; Zapata et al. 2006; Rosero et al. 2016). In some cases (18102$-$1800, 18290$-$0924, 18488+0000, 20293+3952), the dominant source in the field is a bright, extended centimeter source which also dominates the mid-IR emission and appears as an extended red feature in the Spitzer images. Close to these extended red sources, there are compact green objects. In these four cases, the masers are projected against the compact objects --- not the more-extended (and presumably more-evolved) centimeter sources. Clearly it would be misleading to associate these masers with the HII regions, even though the latter are present in the field.

\section{Summary and conclusions \label{sec:summ}}

This paper (Paper IV) presents VLA observations of the 44 GHz Class I methanol maser transition toward a sample of 56 high-mass protostellar objects selected from the catalog of Sridharan et al. (2002). Our main conclusions can be summarized as follows:

\begin{enumerate}

\item We detect the 44 GHz transition of Class I methanol maser emission in 24 of the 56 fields observed (a 43\% detection rate). 

\item We find a total of 83 maser components with linewidths ranging from 0.17 km s$^{-1}$ to 3.3 km s$^{-1}$. A histogram of the linewidths shows a flat distribution without a clear peak; the median linewidth is 1.1 km s$^{-1}$.

\item The maser velocities relative to the host molecular clouds range from $-$2.5 to +3.1 km s$^{-1}$. The distribution of these relative velocities peaks near zero but we note a possible small redshift asymmetry.

\item The Spitzer/IRAC images together with VLA centimeter continuum emission reveal multiple sources in different evolutionary stages in each field. The majority of the 44 GHz masers appear to favor the younger sources in each region. The more notorious cases are 18102$-$1800, 18290$-$0924, 18488+0000 and 20293+3952 where the 44 GHz masers are preferentially located toward the more compact and presumably younger object in the field.

\item We report a possible instance of a 44 GHz maser associated with a low-mass protostar in the region IRAS 20293+3952. If confirmed, this region will be the fifth known star-forming region that hosts Class I masers associated with low-mass protostars.

\item Three of the 24 fields with maser emission were catalogued as EGOs by Cyganowski et al. (2008; 18182$-$1433, 18264$-$1152 and 18566+0408). We suggest that at least nine more fields host objects with green excess and morphology similar to EGOs (05358+3543, 18151$-$1208, 18247$-$1147, 18290$-$0924, 18345$-$0641, 18488+0000, 18517+0437, 20126$+$4104 and 20293$+$3952). The spatial coincidence of the 44 GHz masers and the shocked molecular gas traced by the green excess supports the idea that these masers may arise from molecular outflows.

\item A comparison of our results with previous VLA surveys suggests that IRAS 20126+4104 is a plausible case of maser variability. We detect bright masers (about 20 Jy) in sources 05358+3543 and 18264$-$1152 which were not detected by single-dish observations, suggesting variability.

\end{enumerate}

A final paper (Paper V) will present a complete statistical analysis of the entire survey data comprising around 100 high-mass star forming regions.

\acknowledgments

 We thank the anonymous referee for the useful comments that helped to improve the quality of this paper. We thank Aina Palau and Viviana Rosero for providing radio continuum and line images of several sources. We are grateful to T.K. Sridharan for helping us to understand the differences between the positions reported in their survey relative to the IRAS catalog and to Luis Felipe Rodr\'iguez for useful discussions. CBRG thankfully aknowledges the financial support of CONACyT. CBRG and SEK acknowledge partial support from UNAM DGAPA grant IN 114514. PH acknowledges support from NSF grant AST-0908901 for this work. SVK acknowledges partial support from RFBR grants no.~10-02-00147 and 15-02-07676. 

\facility{VLA}
\software{AIPS (van Moorsel, Kemball, \& Greisen 1996), DS9} 


\end{document}